\documentclass{aa}

\usepackage{graphicx}
\usepackage{txfonts}
\usepackage{hyperref}
\usepackage[dvipsnames]{xcolor}
\usepackage{threeparttable}
\usepackage{lscape}
\usepackage{longtable}
\usepackage{placeins}
\usepackage{multirow}

\usepackage{orcidlink}  

\begin{document}

   \title{Gamma-ray flares from the jet of the blazar CTA 102 in 2016--2018}


   \titlerunning{Gamma-ray Flares from the Jet of CTA 102}
   \authorrunning{S. Kim et al. 2024}

   \author{Sanghyun Kim\inst{1,2}
   \orcidlink{0000-0001-7556-8504}
          \and
          Sang-Sung Lee\inst{1,2}\thanks{Corresponding author; \href{mailto:sslee@kasi.re.kr}{sslee@kasi.re.kr}}
          \orcidlink{0000-0002-6269-594X}
          \and
          Juan Carlos Algaba\inst{3}
          \orcidlink{0000-0001-6993-1696}
          \and
          Bindu Rani\inst{4,5}
          \orcidlink{0000-0001-5711-084X}
          \and
          Jongho Park\inst{6,7}
          \orcidlink{0000-0001-6558-9053}
          \and 
          Hyeon-Woo Jeong\inst{1,2}
          \orcidlink{0009-0005-7629-8450}
          \and 
          Whee Yeon Cheong\inst{1,2}
          \orcidlink{0009-0002-1871-5824}
          \and 
          Filippo D’Ammando\inst{8}
          \orcidlink{0000-0001-7618-7527}
          \and
          Anne L\"ahteenm\"aki\inst{9,10}
          \orcidlink{0000-0002-0393-0647}
          \and 
          Merja Tornikoski\inst{9}
          \orcidlink{0000-0003-1249-6026}
          \and 
          Joni Tammi\inst{9}
          \orcidlink{0000-0002-9164-2695}
          \and 
          Venkatessh Ramakrishnan\inst{9,11}
          \orcidlink{0000-0002-9248-086X}
          \and
          Iv\'an Agudo\inst{12}
          \orcidlink{0000-0002-3777-6182}
          \and
          Carolina Casadio\inst{13,14}      
          \orcidlink{0000-0003-1117-2863}
          \and
          Juan Escudero\inst{12}
          \orcidlink{0000-0002-4131-655X}
          \and
          Antonio Fuentes\inst{12}
          \orcidlink{0000-0002-8773-4933}
          \and
          Efthalia Traianou\inst{12}
          \orcidlink{0000-0002-1209-6500}
          \and
          Ioannis Myserlis\inst{15}
          \orcidlink{0000-0003-3025-9497}
          \and
          Clemens Thum\inst{15}
          }

    \institute{Korea Astronomy and Space Science Institute, 776 Daedeok-daero, Yuseong-gu, Daejeon 34055, Republic of Korea
         \and
             Department of Astronomy and Space Science, University of Science and Technology, 217 Gajeong-ro, Yuseong-gu, Daejeon 34113, Republic of Korea
         \and
             Department of Physics, Faculty of Science, {Universiti Malaya}, 50603 Kuala Lumpur, Malaysia
         \and
             NASA Goddard Space Flight Center, Greenbelt, MD 20771, USA
         \and
             Department of Physics, American University, Washington, DC 20016, USA
         \and
             School of Space Research, Kyung Hee University, 1732, Deogyeong-daero, Giheung-gu, Yongin-si, Gyeonggi-do 17104, Republic of Korea
         \and
             Institute of Astronomy and Astrophysics, Academia Sinica, P.O. Box 23-141, Taipei 10617, Taiwan, R.O.C.
         \and
             INAF – Istituto di Radioastronomia, Via Gobetti 101, I-40129 Bologna, Italy
         \and
             Aalto University Mets\"ahovi Radio Observatory, Mets\"ahovintie 114, 02540 Kylm\"al\"a, Finland
         \and
             Aalto University Department of Electronics and Nanoengineering, P.O. BOX 15500, FI-00076 AALTO, Finland
         \and
             Finnish Centre for Astronomy with ESO (FINCA), University of Turku, Vesilinnantie 5, 20014 University of Turku, Finland
         \and
             Instituto de Astrof\'isica de Andaluc\'ia-CSIC, Glorieta de la Astronom\'ia, E-18008, Granada, Spain
         \and
             Institute of Astrophysics, Foundation for Research and Technology - Hellas, Voutes, 7110 Heraklion, Greece
         \and    
             Department of Physics, University of Crete, 70013, Heraklion, Greece
         \and
             Institut de Radiostronomie Milim\'etrique, Avenida Divina Pastora, 7, Local 20, E-18012 Granada, Spain
             }      
   \date{Received XXX XX, XXXX; accepted XXX XX, XXXX}


  \abstract
  {CTA 102 is a $\gamma$-ray bright blazar that exhibited multiple flares in observations by the Large Area Telescope on board the \textsl{Fermi Gamma-Ray Space Telescope} during the period of 2016--2018. We present results from the analysis of multi-wavelength light curves aiming at revealing the nature of $\gamma$-ray flares from the relativistic jet in the blazar. We analyse radio, optical, X-ray, and $\gamma$-ray data obtained in a period from 2012 September 29 to 2018 October 8. We identify {six} flares in the $\gamma$-ray light curve, showing a harder-when-brighter-trend in the $\gamma$-ray spectra. We perform a cross-correlation analysis of the multi-wavelength light curves. We find nearly zero time lags between the $\gamma$-ray and optical and X-ray light curves, implying a common spatial origin for the emission in these bands. We find significant correlations between the $\gamma$-ray and radio light curves as well as negative/positive time lags with the $\gamma$-ray emission lagging/leading the radio during different flaring periods. The time lags between $\gamma$-ray and radio emission propose the presence of multiple $\gamma$-ray emission sites in the source. As seen in 43~GHz images from the Very Long Baseline Array, two moving disturbances (or shocks) were newly ejected from the radio core. The $\gamma$-ray flares from 2016 to 2017 are temporally coincident with the interaction between a traveling shock and a quasi-stationary one at $\sim$0.1~mas from the core. The other shock is found to emerge from the core nearly simultaneous with the $\gamma$-ray flare in 2018. Our results suggest that the $\gamma$-ray flares originated from shock-shock interactions.}

   \keywords{Radiation mechanisms: non-thermal -- Gamma rays: galaxies -- Galaxies: active -- Galaxies: jets -- quasars: individual: CTA 102}

   \maketitle
%

\section{Introduction} \label{sec: intro}

Blazars are among the brightest sources in the universe \citep{Angel+1980, Hovatta+2019}. The preferential orientation of blazar jets is close to our line of sight \citep[for instance, $< 10$~deg;][]{Readhead+1978, Blandford+1979, Hovatta+2009, Liodakis+2018a}, resulting in rapid variability, relativistically-boosted emission, and superluminal jet motions \citep{Blandford+1977, Dondi+1995, Fuhrmann+2014}. Blazars are traditionally classified as flat-spectrum radio quasars (FSRQs) and BL Lac objects (BL Lacs). FSRQs are high-power blazars with prominent broad emission lines in their optical spectra, whereas BL Lacs are low-power ones with weak or absent optical emission lines \citep{Urry+1995, Blandford+2019}.

In more than 15~years of observations of the extragalactic sky with the Large Area Telescope on board the \textsl{Fermi Gamma-Ray Space Telescope} \citep{Atwood+2009}, blazars constitute the largest fraction of the $\gamma$-ray sources in the extragalactic sky \citep[e.g.][]{Abdo+2010b, Nolan+2012, Acero+2015, Abdollahi+2020}. Many blazars exhibit extreme flaring activity in $\gamma$-rays, the causes of which are still a matter of debate \citep[e.g.][]{Tavecchio+1998, Ghisellini+2005, Sikora+2009, MacDonald+2015}. The $\gamma$-ray emission from relativistic jets is generally attributed to inverse Compton (IC) scattering of low-energy photons by synchrotron emitting electrons in the jet \citep{Boettcher+2012, Madejski+2016, Hovatta+2019}. The target photons may come from synchrotron radiation by the same population of relativistic electrons or from radiation fields external to the jet (e.g. the broad-line region (BLR), dusty torus, etc.).

For the former case, the IC scattering is referred to as synchrotron self-Compton \cite[SSC, e.g.][]{Maraschi+1992} and for the latter one, it is called external Compton \citep[EC, e.g.][]{Dermer+1992}. $\gamma$-ray flares are often accompanied by multi-wavelength {flux variations} in radio, optical, or X-rays. Significant correlations between $\gamma$-ray and radio {light curves} have been reported by many studies \citep[e.g.][]{Pushkarev+2010, Agudo+2011, Rani+2013, Fuhrmann+2014, Max-Moerbeck+2014a, Liodakis+2018b}. These correlations, along with Very Long Baseline Interferometry (VLBI) observations conducted at angular scales of milliarcseconds (mas), suggest that $\gamma$-ray flares most likely occur within parsec-scale jets \citep[e.g.][]{Jorstad+2010, Ramakrishnan+2014, Lisakov+2017, Algaba+2018, KimDW+2022}.

CTA~102 ($z = 1.037$) is a $\gamma$-ray bright FSRQ, {hosting a supermassive black hole (SMBH) of $8.5 \times 10^{8}$~${\rm M}_{\odot}$ \citep{Zamaninasab+2014}.} This source exhibits multi-wavelength variability from radio to $\gamma$-rays \citep[e.g.][]{Casadio+2015, Raiteri+2017, D'Ammando+2019}. Previous studies investigated the variability of the source in the 15~GHz radio band in 2012--2018 observations, finding {variability} on timescales of the order of months \citep{KimSH+2022}.

The \textsl{Fermi}-LAT observations of CTA~102 have detected $\gamma$-ray activity.
{It has been suggested that the 2012 $\gamma$-ray flare occurred a few pc from the black hole and coincided with the ejection of a new superluminal jet component from the 43~GHz core \citep{Casadio+2015}. Additionally, the source exhibited similar behavior in the optical and $\gamma$-ray bands during the $\gamma$-ray flare, which is consistent with the framework of the SSC process \citep{Larionov+2016}.}

Many studies investigated $\gamma$-ray flares from CTA~102 during 2016--2017 \citep{Zacharias+2017, Li+2018, Shukla+2018, Kaur+2018, Gasparyan+2018, Prince+2018, Casadio+2019, D'Ammando+2019, Sahakyan+2020, Geng+2022, Sahakyan+2022}. {A $\gamma$-ray flare observed in 2016 January was localized to 5.7--16.7~pc from the central engine \citep{Li+2018}.} From late 2016 to early 2017, the source was in a prolonged period of activity, reaching its highest $\gamma$-ray flux in 2016 December, marking one of the brightest $\gamma$-ray flares in blazars \citep{Ciprini+2016}. Multi-wavelength studies of the 2016 December $\gamma$-ray flare found coincident optical and X-ray flares, suggesting a common origin for the emission in these bands \citep{Kaur+2018, Gasparyan+2018, D'Ammando+2019}. The {multi-wavelength flux variations during the flare} have been interpreted as arising due to inhomogeneities in a curved jet that cause different emission regions to change their orientations and hence their Doppler factors \citep{Raiteri+2017, D'Ammando+2019}. {One scenario suggests that the flare resulted from the ablation of a gas cloud passing through the jet \citep{Zacharias+2017, Zacharias+2019}.} Moreover, investigations of the spectral energy distribution (SED) of CTA 102 during the flare demonstrate that the flare can be explained by leptonic processes \citep{Prince+2018, Gasparyan+2018, Sahakyan+2020, Sahakyan+2022}. Alternatively, \citet{Casadio+2019} explain the flare during 2016--2017 as the interaction between a new jet component and a recollimation shock at $\sim$0.1~mas from the core. Magnetic reconnection has also been proposed as a possible physical mechanism due to the extreme {$\gamma$-ray} variability of the source and the proximity ($\sim$ a few pc) of the flaring region to the central black hole \citep{Geng+2022}.

Additional strong $\gamma$-ray flares were observed from CTA~102 in 2018 \citep[e.g.][]{Geng+2022, Sahakyan+2022}. \citet{Geng+2022} used $\gamma$-ray flaring timescales, the bulk Lorentz factor, and the molecular torus luminosity to calculate an upper limit on the distance between the flaring region and the central black hole of 7.22~pc. The $\gamma$-ray emission in 2018 is well described by a combination of the SSC and EC models \citep{Sahakyan+2022}. However, while several studies explored the 2018 $\gamma$-ray flares, none searched for correlations between $\gamma$-ray and radio {activity} \citep[e.g.][]{Geng+2022, Sahakyan+2022}. We note that no such correlations were found for the major $\gamma$-ray flares that occurred in 2012--2017 \citep[e.g.][]{Casadio+2015, D'Ammando+2019}. For the period of 2018, \citet{KimSH+2022} identified radio {activity}. Our study will be the first to report a correlation between $\gamma$-ray and radio {activity during this period.} Moreover, a correlation study between $\gamma$-ray and radio light curves during flaring activity is crucial to constrain the locations of the $\gamma$-ray regions, which may give us useful information about the nature of $\gamma$-ray flares \citep[e.g.][]{Fuhrmann+2014, Max-Moerbeck+2014b, Algaba+2018, KimDW+2022}.

This study aims at understanding the {physical mechanisms for generating $\gamma$-ray flares and the locations of the emission regions} in the jet of CTA~102 during the period {2016}--2018.
We present results from this work, including multi-wavelength correlations with respect to the $\gamma$-ray light curve and kinematics of the radio jet related to $\gamma$-ray flares.
We provide the first results of examining a correlation between $\gamma$-ray and radio flares during 2017--2018.
In Sect.~\ref{sec: obs&data}, we describe the observations and data {from} radio to $\gamma$-rays. Multi-wavelength light curves are analysed statistically in Sect.~\ref{sec: mwlc}.
We present the cross-correlation results in Sect.~\ref{sec: DCF}.
In Sect.~\ref{sec: jet_kinematics}, we present the pc-scale jet kinematics of CTA~102 at 43~GHz. We discuss the physical {origins} of $\gamma$-ray flares based on our results in Sect.~\ref{sec: discussion}. A summary of our results is given in Sect.~\ref{sec: summary}. Throughout this paper, we adopt a flat ${\rm \Lambda}$CDM cosmology with $\Omega_{\rm m} = 0.315$, $\Omega_{\rm \Lambda} = 0.685$, and $H_0 = 67.4$~${\rm km~s^{-1}~{Mpc}^{-1}}$ \citep{PlanckCollaboration+2020}. The luminosity distance at {the} redshift of the source (i.e. $z = 1.037$) is $D_{\rm L} = 7112.9$~Mpc and {the} linear scale is 8.311~pc~${\rm mas}^{-1}$, calculated with the Cosmology Calculator\footnote{\url{https://astro.ucla.edu/~wright/CosmoCalc.html}} \citep{Wright+2006}. A proper motion of 1~mas~${\rm yr}^{-1}$ corresponds to {55.2c}.

\section{Observations and data} \label{sec: obs&data}

\subsection{Radio data}
We {used} the multi-wavelength radio data from \citet{KimSH+2022} as well as {observations from the Submillimeter Array (SMA) taken} at 340~GHz \citep[for details and references {for} the data, see][]{KimSH+2022}.
The multi-wavelength data {cover} a period from {2012 November 20 to 2018 September 23 (modified Julian date, MJD 56251--58384)} with a frequency range of {15--343}~GHz.
They are complemented by {data from} the Owens Valley Radio Observatory (OVRO) 40-m radio telescope at 15 GHz \citep{Richards+2011}, the Metsähovi 13.7-m radio telescope at 37~GHz \citep{Teraesranta+1998}, the IRAM 30-m {telescope} at 86 and 230~GHz, the SMA at 230~GHz \citep{Gurwell+2007}, and the Atacama Large millimeter/submillimeter Array (ALMA) at 91, 233, and 343~GHz.
{The 86/230~GHz IRAM data were obtained under the POLAMI (Polarimetric Monitoring of AGN at Millimetre Wavelengths) IRAM Large Program\footnote{\url{http://polami.iaa.es}} \citep{Agudo+2018a, Agudo+2018b, Thum+2018}.}
We note that {in combining data} at similar frequencies, we used the 89~GHz (IRAM + ALMA), 230~GHz (IRAM + SMA + ALMA), and 343~GHz (SMA + ALMA)~GHz data for {the analyses discussed in} this paper.
More detailed descriptions of the multi-wavelength data can be found in \citet{KimSH+2022}.

The VLBA-BU-BLAZAR monitoring program\footnote{\url{https://www.bu.edu/blazars/VLBAproject.html}} has {observed $\gamma$-ray bright blazars}{,} including CTA~102{,} with the Very Long Baseline Array (VLBA) at 43~GHz. 
We used the calibrated VLBA data\footnote{\url{https://www.bu.edu/blazars/VLBA_GLAST/cta102.html}} observed from 2016 January 1 to 2018 August 26 (MJD 57388--58356){, with a total of 27 epochs}. Details of the data calibration are described in \citet{Jorstad+2017}. 
The imaging results using the \emph{CLEAN} task in the interferometric imaging software \texttt{DIFMAP} \citep{Shepherd+1997} are provided on the VLBA-BU-BLAZAR monitoring program website. 
Further analysis (e.g. Gaussian model-fitting) has been performed on the data {as} described in Sect.~\ref{sec: jet_kinematics}.

\subsection{Optical data}
We used {data taken in the \textsl{V} and \textsl{R} bands by the optical spectropolarimetric monitoring program of the Steward Observatory at} the University of Arizona.
This program utilizes either the {2.3-m} Bok Telescope on Kitt Peak {or the 1.54-m Kuiper Telescope on Mt. Bigelow \citep[see][for a more detailed description]{Smith+2009}.}
The data {span} a period from {2012 October 11 to 2018 July 7} (MJD 56211–58306).
The reduced photometry data are publicly {available.\footnote{\url{http://james.as.arizona.edu/~psmith/Fermi}}}
{The magnitudes} of the optical data have been corrected for Galactic extinction with values in the NASA/IPAC Extragalactic Database {(NED).\footnote{\url{http://ned.ipac.caltech.edu/}}}
The effective wavelengths for {the \textsl{V} and \textsl{R} bands} are 0.55 and 0.64~${\rm \mu m}$, respectively \citep{Schlafly&Finkbeiner+2011}. 
We converted the corrected magnitudes {into flux densities} (in units of mJy) using zero-magnitude flux densities \citep{Mead+1990}.

\subsection{Hard X-ray data}
The \textsl{Neil Gehrels Swift Observatory} \citep{Gehrels+2004} detected the source during 157 months from 2012 September 29 to 2017 December 30 (MJD~56200--58118).
The observations were performed with the Burst Alert Telescope \citep[BAT;][]{Barthelmy+2005} {and will be included in the upcoming \textsl{Swift}-BAT 157-Month Hard X-ray Survey (Lien et al. in {prep}.).\footnote{\url{https://swift.gsfc.nasa.gov/results/bs157mon/}}}
We obtained the Crab-weighted monthly binned light curve at 14--195~keV.
We flagged the data when the count rate is below zero.

\subsection{\texorpdfstring{$\gamma$}{Lg}-ray data}

We have analysed the $\gamma$-ray data from the \textsl{Fermi}-LAT during the period from 2012 September 29 to 2018 October 8 (MJD 56200--58400), following the standard unbinned likelihood procedure. We performed the analysis with the \texttt{SCIENCETOOLS}\footnote{\url{https://fermi.gsfc.nasa.gov/ssc/data/analysis/}} software package version v11r5p3. We selected the SOURCE class events (\texttt{evclass = 128, evtype = 3}) at 100~MeV--300~GeV.
{To ensure that we used} good quality data, we used the filter parameters ${\rm DATA\_Qual>0}~\&\&~{\rm LAT\_CONFIG==1}$.
We applied the maximum zenith angle of 90~$\deg$ to reduce contamination from Earth limb $\gamma$-rays.
We defined a Region of Interest (ROI) of 15~$\deg$ radius centred at the location of CTA~102. We used \texttt{P8R3\_SOURCE\_V1} as the instrument response function. In order to model the background, we also used the {Galactic} diffuse emission model gll\_iem\_v07 and the isotropic background model iso\_P8R3\_SOURCE\_V3\_v1.

The spectral analysis was performed using the Science tool \texttt{gtlike}.
The source model included all sources from the fourth \textsl{Fermi} Large Area Telescope catalog \citep[4FGL;][]{Abdollahi+2020} that appeared within 30~deg of the centre of the ROI.
{For 4FGL sources, we parameterized their spectra} using a power law (PL), a log-parabola (LP), or a PL with a super-exponential cut-off {\citep[see e.g.][for detailed equations]{Prince+2018}.}
Model parameters for the sources within 10~$\deg$ of the centre of {the} ROI and for those whose detection {significances are} {$\geq 15$} were set free.
For CTA~102, the PL model was used, in the form of ${\rm d}N(E) / {\rm d}E = N_0{(E/E_0)}^{-\Gamma}$, where $N_0$ is {the} prefactor, $\Gamma$ is {the} photon index, and $E_0$ is {the} pivot energy.
We {left the prefactor and the photon index free to vary} and used the pivot energy $E_0 = 414$~MeV from 4FGL \citep{Abdollahi+2020}.
The normalisation factors of variable sources {(those with variability indexes $\geq$ 18.48 in the 4FGL catalog) that were within 10~deg of CTA~102 were also left free to vary.}
For the other sources, the normalisation factor and the spectral shape parameters were fixed. 
We evaluated the {significances of the $\gamma$-ray signals from the sources using a maximum-likelihood analysis and setting a test statistic (TS) value of 10 as the detection limit.}

\begin{figure*}[htb!]
\centering
\includegraphics[scale=0.8]{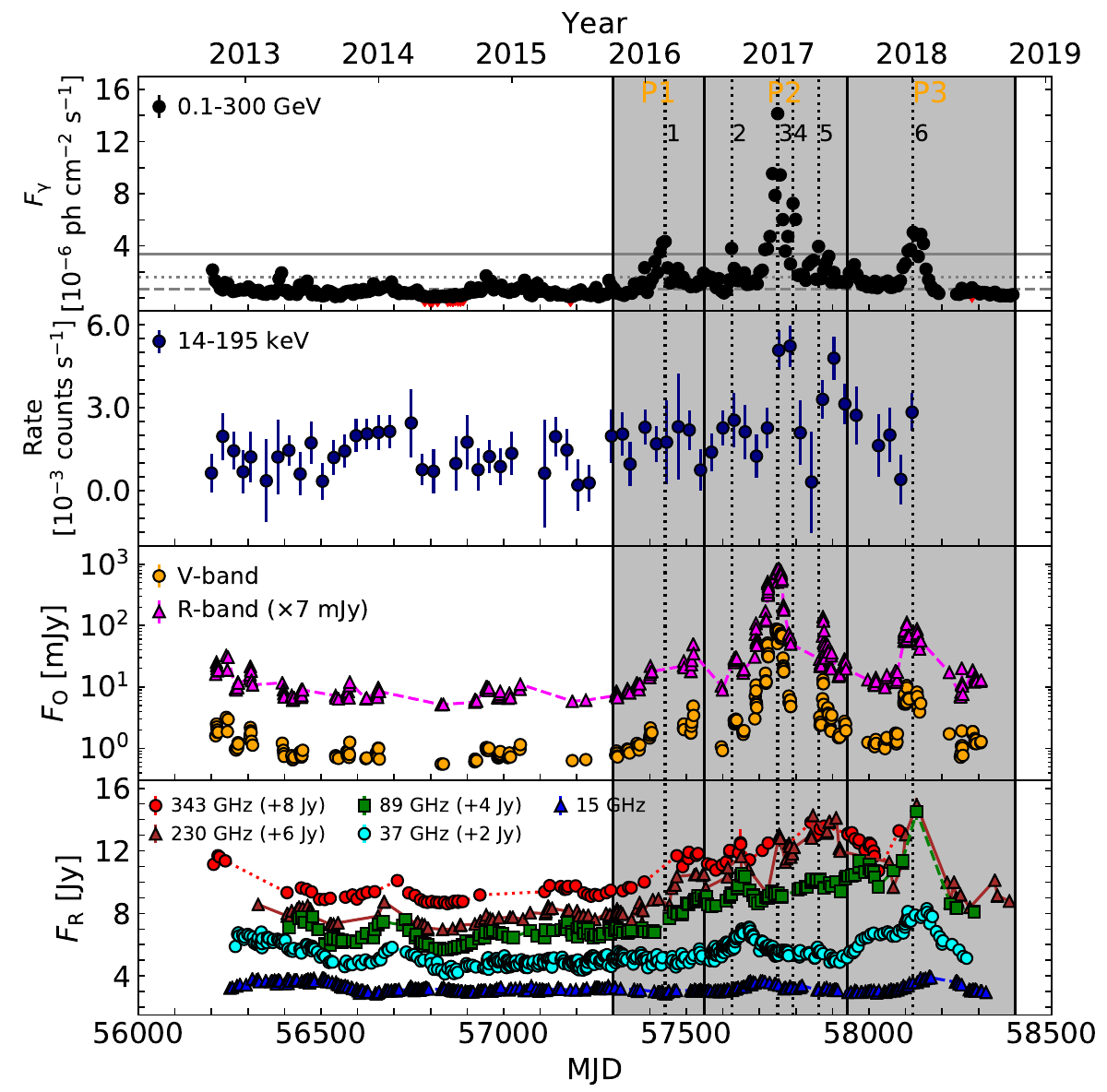}
\caption{Multi-wavelength light curves from CTA~102. The entire period is from 2012 September 30 to 2018 October 9 (MJD 56200--58400). From top to bottom, the panels show light curves in $\gamma$-rays, X-rays, optical, and radio with different colours and symbols indicating different frequencies. The red triangles in the $\gamma$-ray light curve indicate upper limits on the flux density. The optical and radio light curves are displayed with arbitrary offsets for clarity. The grey {horizontal} lines for the $\gamma$-ray light curve denote the flux density threshold (see the text for details). The black dotted lines indicate the peaks of the $\gamma$-ray flares (1--6). The grey shaded areas mark {three periods (P1--3) for the cross-correlation analysis.} 
}
\label{fig: light curves}
\end{figure*}

We performed the first maximum-likelihood analysis over the whole period.
The second maximum-likelihood analysis was performed after removing the sources with TS $<$ 10. 
Applying this result to the source model, we generated a weekly binned light curve for the entire period in this work.
One of the solar arrays on the \textsl{Fermi}-LAT experienced an anomaly on 2018 March 16.\footnote{\url{https://fermi.gsfc.nasa.gov/ssc/observations/types/post_anomaly/}}
We did not make use of the data from mid-March to early April 2018.
The fit results {for integrating over the whole period yield a TS value of 340798 in the 100~MeV--300~GeV energy range. The integrated average flux is $(129.4~\pm~0.5) \times 10^{-8}~{\rm ph~cm^{-2}~s^{-1}}$, and the photon index is $2.239 \pm 0.003$.}

\section{Multi-wavelength light curves}  \label{sec: mwlc}

Fig.~\ref{fig: light curves} presents multi-wavelength light curves from CTA 102 in the {$\gamma$-ray, X-ray, optical, and radio} bands. The light curves cover a time range from 2012 September 30 to 2018 October 9 (MJD~56200--58400). We define the $\gamma$-ray brightness state (i.e. a quiescent state and an active state) of the source based on the weighted mean flux of $\left<F_{\omega}\right> = 0.7 \times 10^{-6}$~${\rm ph~cm^{-2}~s^{-1}}$ and the weighted standard deviation of $\sigma_{\omega} = 0.9 \times 10^{-6}$~${\rm ph~cm^{-2}~s^{-1}}$ \citep{Williamson+2014}. The source is quiescent with flux densities at the level of $F_{\rm \gamma} < \left<F_{\omega}\right>$, while active with those at the level of $F_{\rm \gamma} > \left<F_{\omega}\right> + \sigma_{\omega}$. We regard a threshold as $\left<F_{\omega}\right> + 3\sigma_{\omega} = 3.4 \times 10^{-6}$~${\rm ph~cm^{-2}~s^{-1}}$ to define a flaring state. Until late 2015, the $\gamma$-ray brightness of the source was quiescent with flux densities at the level of $F_{\rm \gamma} \leq 1.6 \times 10^{-6}$~${\rm ph~cm^{-2}~s^{-1}}$. After 2016, the source became active in $\gamma$-rays.

{For the period from 2015 October 5 to 2018 October 9 (MJD~57300--58400), we identified six $\gamma$-ray flares}, labelled (1--6) in Fig.~\ref{fig: light curves}. Flare~3 is the brightest $\gamma$-ray flare, reaching a peak of $(1.415 \pm 0.018)\times 10^{-5}$~${\rm ph~cm^{-2}~s^{-1}}$ on 2016 December 28 (MJD~57750). {\citet{D'Ammando+2019} studied this source for the period of 2013 January 1--2017 February 9 (MJD~56293--57793), which overlaps with the period studied here. Three of the flares identified in this work (Flares~1, 2, and 3) were also identified by that group (denoted Flares~IV, V, and VI in their work).}

\begin{figure}[htb!]
\centering
\includegraphics[scale=0.5]{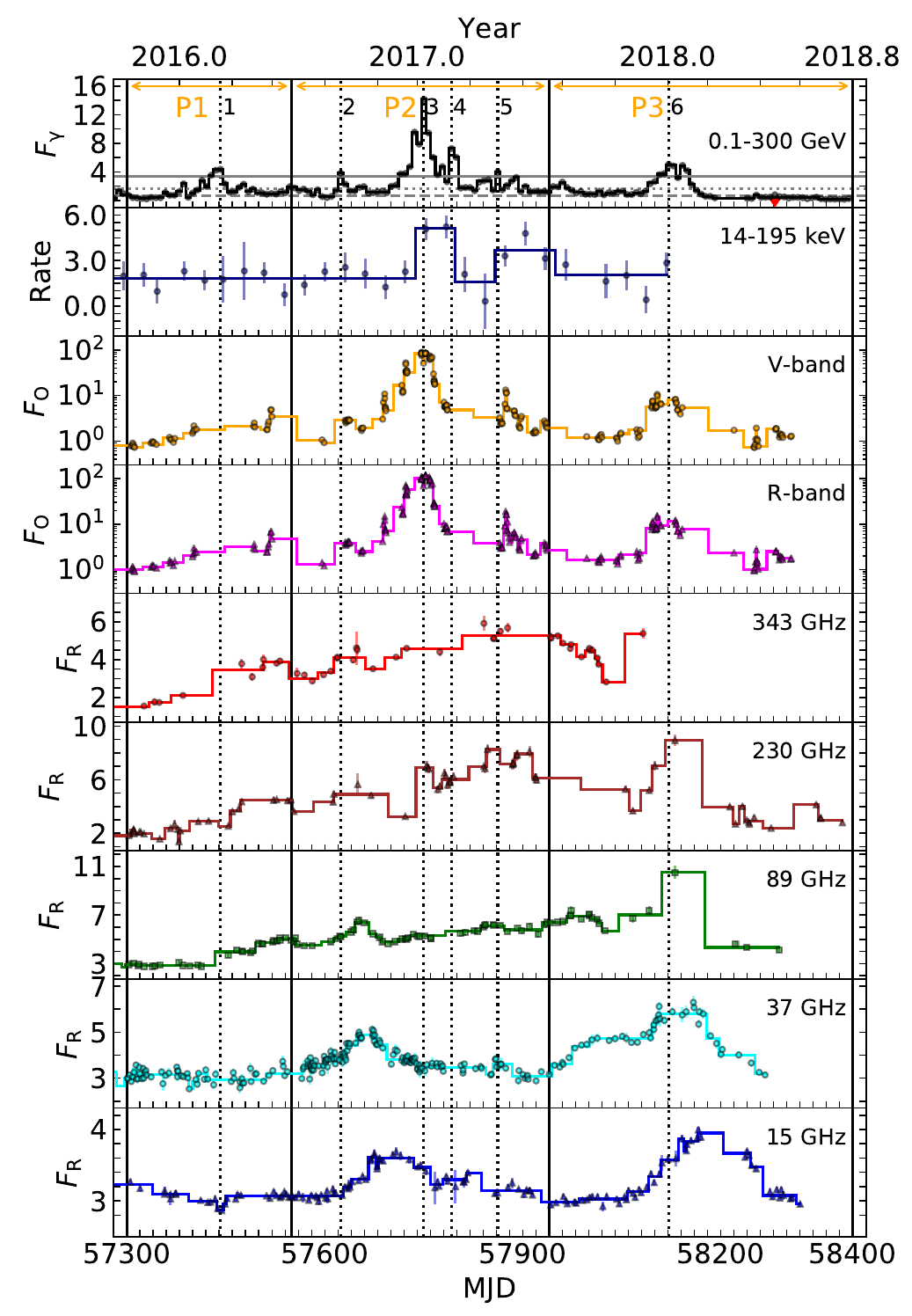}
\caption{Bayesian block representation of the multi-wavelength light curves during the flaring period. The time range of the period is from 2015 October 5 to 2018 October 9 (MJD~57300--58400). The units of the flux densities and vertical lines are the same as in Fig.~\ref{fig: light curves}. The horizontal arrows indicate the time ranges of flaring periods (P1--3).}
\label{fig: light curves+FP+BB}
\end{figure}

We employed a Bayesian block analysis \citep{Scargle+2013} to {identify the local maxima of multi-wavelength light curves.} This method divides the light curves into blocks so that the flux measurements within each block are statistically consistent with a constant value. In this analysis, we set a false positive rate of 0.01 (the probability of falsely reporting the detection of a change point) for all energy bands. We identify a local maximum if the flux density of the block exceeds those of both previous and subsequent blocks. Fig.~\ref{fig: light curves+FP+BB} presents the Bayesian block representation of the light curves in the $\gamma$-ray, X-ray, optical, and radio bands, zooming in on the period that includes the $\gamma$-ray flares (see above).
{Our analysis revealed the following number of local maxima: 24 in the $\gamma$-ray light curve, 2 in the X-ray light curve, 31 in the optical \textsl{V}-band light curve, 21 in the optical \textsl{R}-band light curve, and 4–-11 in the radio light curves (4 at 343~GHz, 5 at 15/37~GHz, 6 at 89~GHz, and 11 at 230~GHz).}
We note {a larger number of local maxima in the optical and $\gamma$-ray light curves compared to the radio band, suggesting more rapid variability at higher energies.
The small number of local maxima in X-rays is mainly related to the relatively low sensitivity of \textsl{Swift}-BAT.}
\citet{KimSH+2022} have reported that the rise time of radio flares in the jet of the source is $< 250$~days.

   \begin{table}[htb!]
      \footnotesize
      \caption[]{{$\gamma$-ray {flares} in 2016--2018}}
         \label{tab: gamma-ray flaring periods}
      \centering
        \begin{tabular}{lclcc}
            \hline
            \hline
            \noalign{\smallskip}
            Period & Duration & ID & Peak  & Flux density \\
            {}     & (MJD)    & & (MJD) & ($\times 10^{-6}~{\rm ph~cm^{-2}~s^{-1}}$) \\
            \noalign{\smallskip}
            \hline
            \noalign{\smallskip}
            P1                     & 57300--57550  & 1 & 57442 & $4.34 \pm 0.10$ \\
            {P2}            & {57550--57940}   & 2 & 57624 & $3.81 \pm 0.13$ \\
                                   &                & 3 & 57750 & $14.15 \pm 0.18$ \\
                                   &                & 4 & 57792 & $7.26 \pm 0.17$ \\
                                   &                & 5 & 57862 & $3.96 \pm 0.12$ \\
            P3                     & 57940--58400   & 6 & 58121 & $5.05 \pm 0.14$ \\
            \noalign{\smallskip}
            \hline
        \end{tabular}
        \scriptsize
        \tablefoot{{The columns are (1) identification of {a period for the cross-correlation analysis}, (2) {duration of the period in MJD}, (3) identification of a $\gamma$-ray flare, (4) MJD of the peak during the period, and (5) peak flux density of the period in the unit of ${\rm ph~cm^{-2}~s^{-1}}$.}}
    \end{table}

{To explore the cross-correlations between $\gamma$-ray and lower-energy light curves in Sect.~\ref{sec: DCF}, we split the period into three parts, i.e. P1, P2, and P3, based on the Bayesian block analysis results.
The period P2 is defined to include the strongest $\gamma$-ray flare (Flare~3) and its neighboring flares (Flares~2, 4, and 5), as well as the most prominent local maxima in the X-ray and optical light curves.
This period extends from MJD~57550 to MJD~57940.
The beginning and ending epochs of the period are aligned with the local minima identified in the Bayesian block analysis of the well-sampled 15~GHz light curve.
The periods preceding and following P2 within the overall flaring period were determined as P1 (MJD~57300--57550) and P3 (MJD~57940--58400), respectively.}
Table~\ref{tab: gamma-ray flaring periods} provides the duration of the periods and the peak information of the $\gamma$-ray flares.

\begin{figure}[htb!]
\includegraphics[scale=0.55]{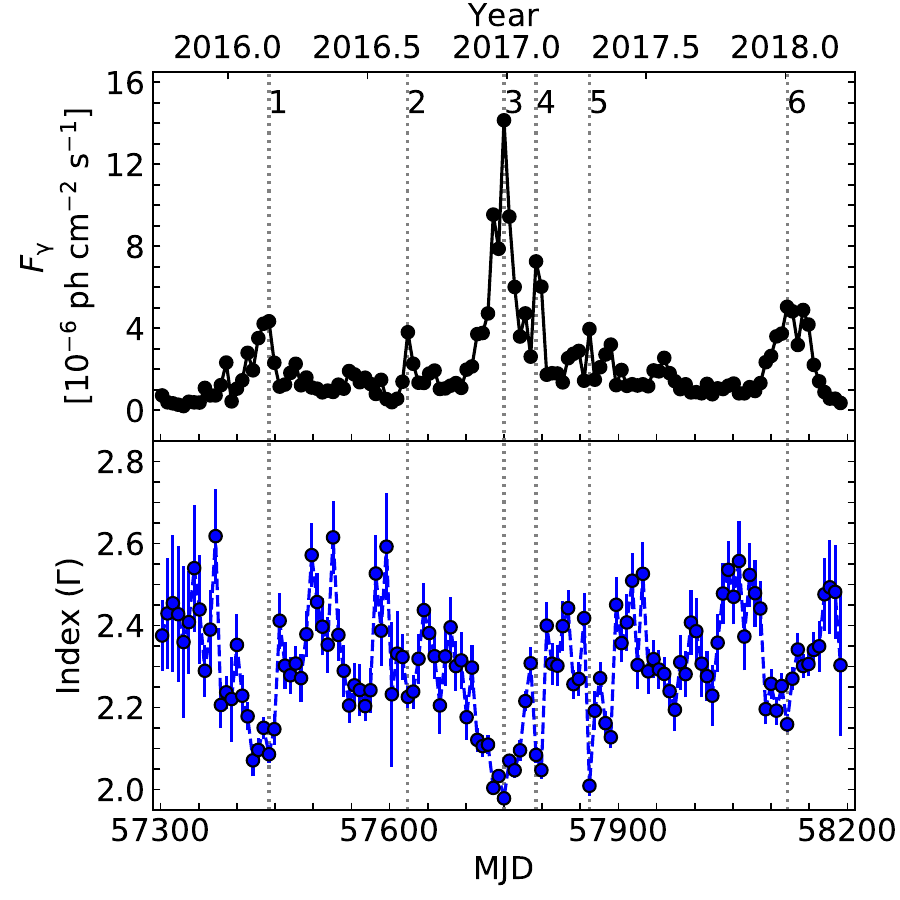}
\centering
\caption{$\gamma$-ray light curve (top) and time {series for the} $\gamma$-ray photon index (bottom).
Black dotted lines (1--6) indicate the peaks of the $\gamma$-ray flares. }
\label{fig: gamma lc & index}
\end{figure}

We searched for variations in the $\gamma$-ray photon index over the flaring period divided into seven-day time bins. Fig.~\ref{fig: gamma lc & index} presents the $\gamma$-ray light curve in the upper panel and the $\gamma$-ray photon index in the lower panel.  The black dotted lines (1--6) indicate $\gamma$-ray flares that are the same as in Fig.~\ref{fig: light curves}. The $\gamma$-ray photon index decreases when the source becomes brighter, indicative of spectral hardening (also known as the harder-when-brighter trend). We note that the $\gamma$-ray spectrum also becomes harder when small flux enhancements (e.g. between Flares~1 and~2 or between Flares~5 and~6) are observed below the $\gamma$-ray flaring threshold.

\begin{figure}[htb!]
\includegraphics[scale=0.6]{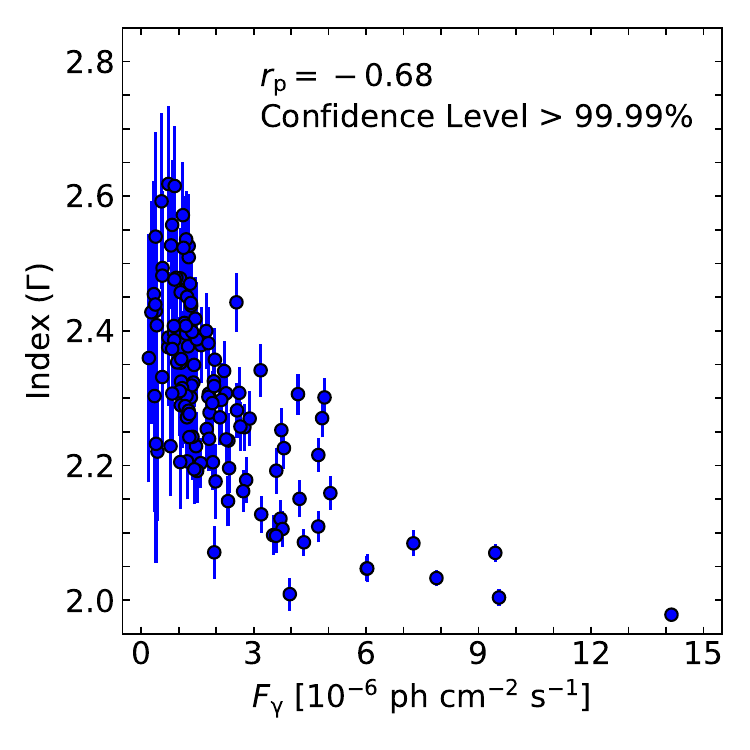}
\centering
\caption{$\gamma$-ray photon index ($\Gamma$) versus $\gamma$-ray flux density ($F_\gamma$). The Pearson correlation coefficient $r_{\rm p}$ is presented with {the} confidence level.}  \label{fig: gamma flux vs index}
\end{figure}

Fig.~\ref{fig: gamma flux vs index} displays the $\gamma$-ray flux density versus the $\gamma$-ray photon index over the flaring period. As in Fig.~\ref{fig: gamma lc & index}, a clear harder-when-brighter trend is seen in Fig.~\ref{fig: gamma flux vs index}. The $\gamma$-ray photon index varies between 1.98 and 2.62 with an average of $2.31 \pm 0.01$. Consistent results were obtained by \citet{D'Ammando+2019} who studied variations in the $\gamma$-ray photon index occurring over a time period of 2013 January 1 to 2017 February 9 (MJD~56293--57793) {divided into} 30-day time bins. Assuming the LP model, {they found a spectral slope that varied} between 1.88 and 2.97 with an average of $2.30 \pm 0.09$. We performed a Pearson correlation analysis to evaluate the correlation between the $\gamma$-ray flux density and the $\gamma$-ray photon index. In quantifying the correlation between photon index and flux density, we calculated the Pearson correlation coefficient, obtaining a value of $r_{\rm p} = -0.68$ for a confidence level greater than 99.99\%. The negative value for $r_{\rm p}$ suggests an anti-correlation. Therefore, we conclude that the harder-when-brighter trend is seen in the $\gamma$-ray band \citep[see also][]{Algaba+2018, Rani+2018, Prince+2020}.

\section{Correlation analysis}   \label{sec: DCF}

\subsection{Method}
We carried out a correlation analysis that is applicable to unevenly-sampled data sets. For this analysis, we calculated the discrete correlation function {(DCF)} in order to test for correlations between the various multi-wavelength light curves and to determine their time lags. {Details on the DCF are described in \citet{Edelson&Krolik+1988}.} A positive (negative) time lag indicates that the higher-frequency time series leads (lags) the lower-frequency time series. For our calculations, we used the \texttt{PYTHON} implementation of the public DCF code developed by \citet{Robertson+2015}.

In order to compute the DCF, there are a few effects that we consider.
\citet{Peterson+2004} found that detrending does not improve the accuracy of estimates of time lags in unevenly sampled time-series data; rather, detrending results in large errors. Thus, we did not detrend our light curves, following \citet{Max-Moerbeck+2014b}. For each frequency pair, we equated the bin width of the time lag ($\Delta t_{\rm bin}$) with the mean cadence ($t_{\rm int}$) of the less frequently sampled light curve. The optimal bin width was determined in a range from $t_{\rm int}$ to $2 t_{\rm int}$. For computing the time lags, we selected a range that covers at least two-thirds of fraction of the light curves \citep[e.g.][]{Liodakis+2018b}.

To determine the significance of the DCF, we simulated artificial light curves with statistical properties, i.e. the power spectral density (PSD) and the probability density function (PDF), similar to the real light curves for each frequency pair. For these simulations, we used the public code developed by \citet{Connolly+2015} that implements the model described in \citet{Emmanoulopoulos+2013}.
{We calculated the PSDs using the normalised Scargle periodogram \citep{Scargle+1982}. In order to fit the periodogram ($P_f$), we used a power-law spectrum, given by $P_f \propto {f}^{-\beta}$, where $\beta$ is the slope of the power-law spectrum \citep{Uttley+2002, Vaughan+2003}. We performed Monte Carlo simulations of light curves to find the best-fit slopes of the observed power spectra \citep[see][for more details on the method]{Park+2014, Park+2017}. We also calculated the PDFs of the multi-wavelength light curves, finding that the optical and $\gamma$-ray light curves follow gamma distributions, while the radio and X-ray light curves follow log-normal distributions \citep[see][for more details on the method]{Algaba+2018}.}
We then calculated the DCFs for a large sample of simulated light curves, obtaining 10,000 DCFs which we used to estimate 1$\sigma$, 2$\sigma$, and 3$\sigma$ (68$\%$, 95$\%$, and 99.7$\%$, respectively) confidence intervals. 

To measure the uncertainty in a DCF peak (${\rm DCF}_{\rm peak}$) and its corresponding time lag ($\tau_{\rm DCF}$), we used the flux randomization and random subset selection (FRRSS) method \citep{Peterson+1998}. Using this method, we simulated 10,000 artificial light curves randomly sampled from each light curve. Then, we performed the DCF analysis and obtained 10,000 ${\rm DCF}_{\rm peak}$ and ${\tau}_{\rm DCF}$ values for each frequency pair. We fit the ${\rm DCF}_{\rm peak}$ and ${\tau}_{\rm DCF}$ distributions with Gaussian functions with means and standard deviations corresponding to the best-fit values for the parameters and their uncertainties, respectively.

\subsection{Results}    \label{subsec: DCF_results}

\begin{figure}[htb!]
\vspace{3mm}
\includegraphics[scale=0.6]{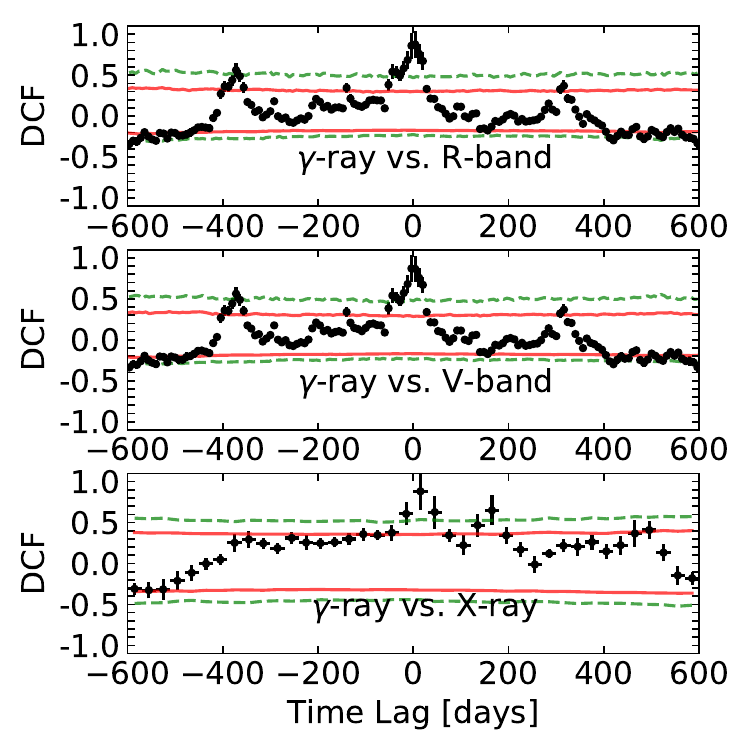}
\centering
\caption{DCFs for $\gamma$-ray versus optical (\textsl{R}-and \textsl{V}-bands) and X-ray over the entire period, from top left to bottom right. The red solid and green dashed lines indicate 2$\sigma$ (95\%) and 3$\sigma$ (99.7\%) confidence intervals, respectively. Positive time lags indicate that the flux density in $\gamma$-rays leads the flux density in lower-energy bands.} \label{fig: DCF_entire-period_opt+hardX-gamma}
\end{figure}

We carried out the cross-correlation analysis over the entire period from 2012 September 30 to 2018 October 9 (MJD 56200--58400) in order to characterize the overall correlation between the $\gamma$-ray light curve and those of the other bands. The analysis reveals significant correlations between the optical/X-ray and $\gamma$-ray light curves above 99.7\% confidence levels, implying a time coincidence and hence co-spatiality of the optical/X-ray and $\gamma$-ray emission regions, given in Fig.~\ref{fig: DCF_entire-period_opt+hardX-gamma}. The 3$\sigma$ correlations come from the cross-match of Flare~3 with {local maxima} in the optical and X-ray bands during {P2}. Cross-correlation analyses during the flaring period (MJD 57300--58400) confirmed a time coincidence between the optical/X-ray and $\gamma$-ray light curves. \citet{D'Ammando+2019} also reported a correlation with a zero time lag between optical/X-ray and $\gamma$-ray light curves, consistent with our results.

   \begin{table*}[htb!]
      \caption[]{{Cross-correlation results for $\gamma$-ray versus multi-wavelength (radio and optical) pairs during {P2 and P3}}}
         \label{tab: DCF_P3}
      \centering
        \begin{tabular}{lcccc}
            \hline
            \hline
            \noalign{\smallskip}
            Frequency Band & \multicolumn{2}{c}{P2}         & \multicolumn{2}{c}{P3}       \\
            {}             & DCF & $\tau_{\rm DCF}$ (day)   & DCF & $\tau_{\rm DCF}$ (day) \\
            \noalign{\smallskip}
            \hline
            \noalign{\smallskip}
            15~GHz          & $0.80 \pm 0.07$  & ${-52 \pm 4}^{\star}$  &   $0.87 \pm 0.04$  &   ${23 \pm 5}^{\star}$    \\
            37~GHz          & $0.86 \pm 0.17$  & ${-86 \pm 7}^{\star}$  &   $0.72 \pm 0.07$  &   $13 \pm 5$  \\
            89~GHz          & $\cdots$         & $\cdots$               &   $0.85 \pm 0.06$  &   ${20 \pm 13}^{\star}$   \\
            230~GHz         & $\cdots$         & $\cdots$               &   $0.91 \pm 0.05$  &   ${4 \pm 11}$ \\
            \textsl{R}-band & $0.96 \pm 0.10$  & ${-1 \pm 3}^{\star}$   &   $0.86 \pm 0.08$  &   ${-12 \pm 5}^{\star}$ \\
            \textsl{V}-band & $0.96 \pm 0.10$  & ${-2 \pm 2}^{\star}$   &   $0.82 \pm 0.08$  &   ${-12 \pm 5}^{\star}$ \\
            \noalign{\smallskip}
            \hline
        \end{tabular}
        \tablefoot{(1) observing frequency, (2) Peak DCFs and uncertainties in P2, (3) time lags ($\tau_{\rm DCF}$) and uncertainties of the DCF computed between the $\gamma$-ray light curve and those in other frequency bands in P2, (4) Peak DCFs and uncertainties in P3, and (5) time lags ($\tau_{\rm DCF}$) and uncertainties of the DCF computed between the $\gamma$-ray light curve and those in other frequency bands in P3. Time lags are marked with a star symbol ($\star$) when the peak DCF is above the 99.7\% confidence level.}
    \end{table*}

\begin{figure}[htb!]
\vspace{3mm}
\includegraphics[scale=0.6]{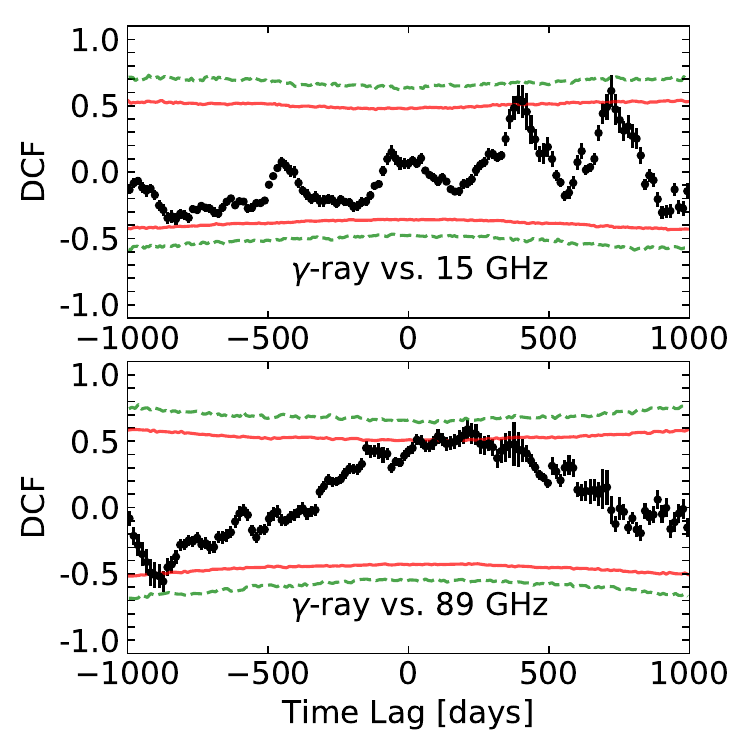}
\centering
\caption{
{DCFs for $\gamma$-ray versus radio (15 and 89~GHz) bands over the entire period. Horizontal lines are the same as in Fig.~\ref{fig: DCF_entire-period_opt+hardX-gamma}.}
} \label{fig: DCF_entire-period_15+89-gamma}
\end{figure}

The $\gamma$-ray light curve exhibited more complicated correlations with the radio light curves than with the optical band. Fig.~\ref{fig: DCF_entire-period_15+89-gamma} presents the DCF results for $\gamma$-ray versus radio (15 and 89~GHz) over the entire period. The analysis reveals double 2$\sigma$ correlations (for 15 and 37 GHz) with positive time lags of a few hundred days (top in Fig.~\ref{fig: DCF_entire-period_15+89-gamma}), as well as 2$\sigma$ correlations (for 89, 230, and 343~GHz) with a broad range of positive time lags (bottom in Fig.~\ref{fig: DCF_entire-period_15+89-gamma}). The former comes from the matching of the 15/37~GHz {local maxima} in P3 with Flare~1 (or Flare~3) based on the peak-to-peak time intervals (see Fig.~\ref{fig: light curves+FP+BB}). The latter occurs because {local maxima in the radio} (i.e. 89--343~GHz) coincide with the $\gamma$-ray flares (e.g. Flare~5 or Flare~6), while others lag behind the $\gamma$-ray flares (e.g. Flare~1 or Flare~2). 

These correlations are consistent with those observed during the flaring period. As noted in the Bayesian block analysis (Sect.~\ref{sec: mwlc}), {the} different variability behaviours between the $\gamma$-ray and radio emissions could affect the correlation {results}. {To further {explore} these correlations}, we investigated cross-correlations between the $\gamma$-ray and lower-energy light curves during {three distinct periods (P1--3)}.

\begin{figure}[htb!]
\vspace{3mm}
\includegraphics[scale=0.43]{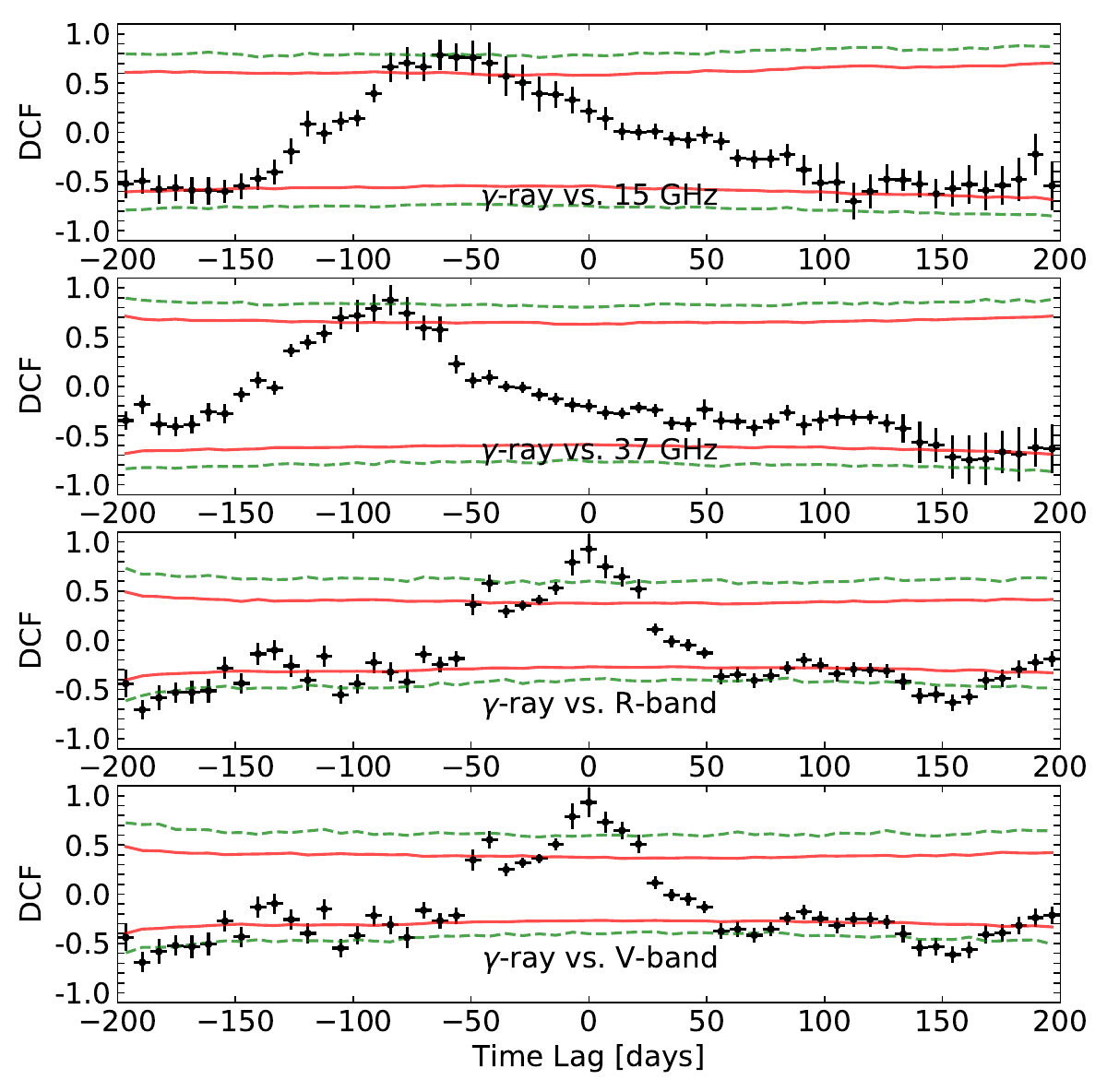}
\centering
\caption{DCFs for $\gamma$-ray versus radio (15/37~GHz) and optical during P2. Horizontal lines are the same as in Fig.~\ref{fig: DCF_entire-period_opt+hardX-gamma}.} \label{fig: DCF_P2}
\end{figure}

We found {no significant correlations for P1. However, significant correlations (above 99.7\% confidence level) were observed between the $\gamma$-ray and radio light curves for P2 and P3.} Fig.~\ref{fig: DCF_P2} presents the observed negative time lags of $< -50$~days in the 3$\sigma$ correlation between the $\gamma$-ray and {15/37~GHz radio} light curves during P2. This correlation is based on matching the 15/37~GHz {local maxima} in P2 with Flare~3, {suggesting} that the $\gamma$-ray emission lagged behind the radio emission. 

\begin{figure}[htb!]
\vspace{3mm}
\includegraphics[scale=0.43]{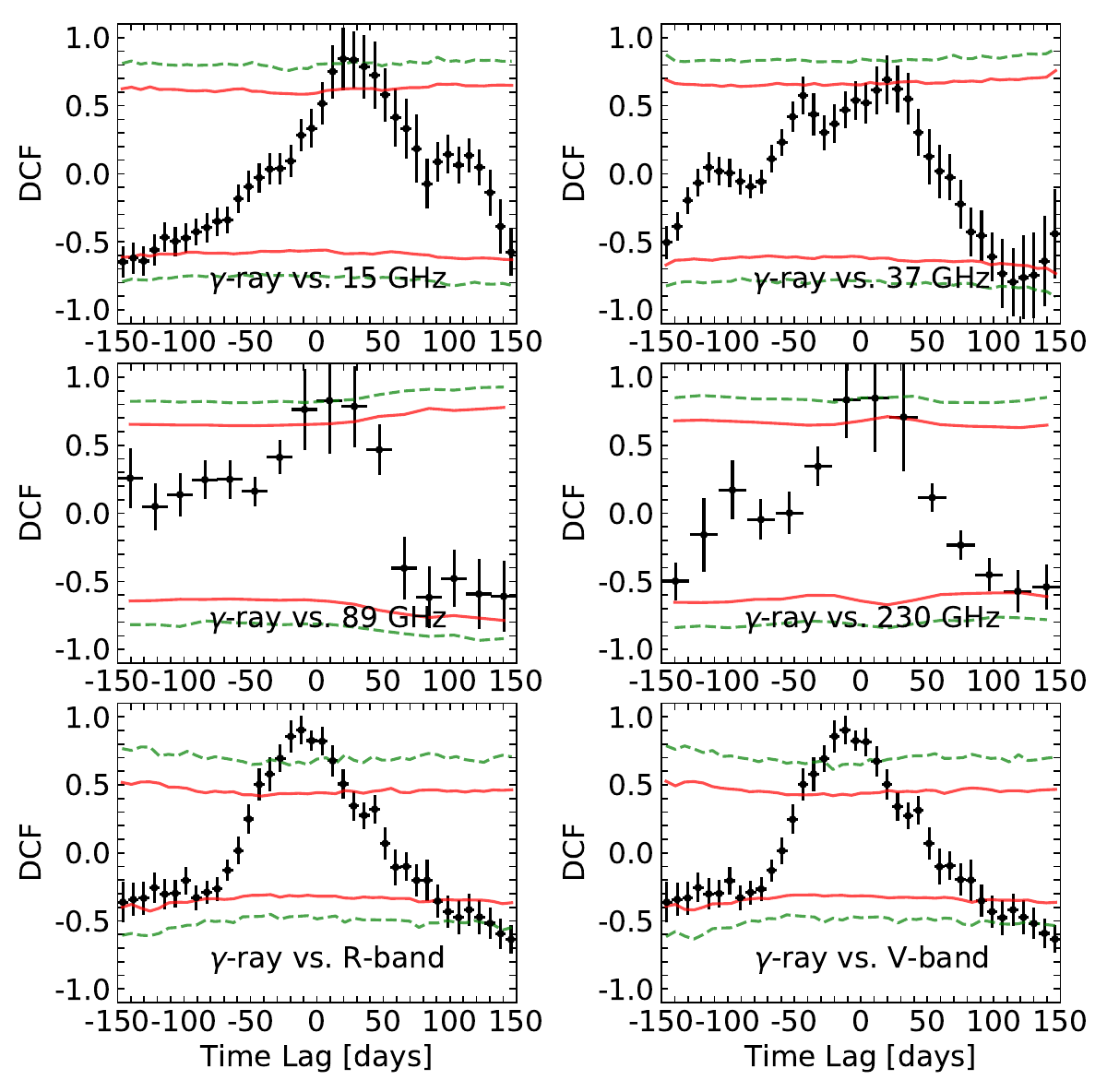}
\centering
\caption{DCFs for $\gamma$-ray versus radio (15, 37, 89, and 230~GHz) and optical (\textsl{R}-and \textsl{V}-bands) in P3 from top left to bottom right. Horizontal lines are the same as in Fig.~\ref{fig: DCF_entire-period_opt+hardX-gamma}.}
\label{fig: DCF_Flare6_radioOpt-gamma-ray}
\end{figure}

Flare~6, the most recent $\gamma$-ray flare in P3, is significantly correlated ($> 99.7$\% confidence levels) with the {local maxima} in the radio band, yielding time lags of less than a month (see Fig.~\ref{fig: DCF_Flare6_radioOpt-gamma-ray}). The positive time lags between the $\gamma$-ray and radio light curves during P3 imply that the $\gamma$-ray emission preceded the radio emission, including the strongest {local maxima} at 15-230 GHz frequencies. {Table~\ref{tab: DCF_P3} summarises the cross-correlation results for P2 and P3.}

In the $\gamma$-radio correlation analysis, we observed both positive and negative time lags, indicating the existence of multiple $\gamma$-ray emission regions in the jet. Positive time lags are a common finding in studies of correlations between radio and $\gamma$-ray {activity} in blazar jets \citep[e.g.][]{Max-Moerbeck+2014b, Ramakrishnan+2015, Algaba+2018, Hodgson+2018, Liodakis+2018b}. \citet{KimSH+2022} examined the synchrotron spectrum of CTA~102 {during P3}, finding that the synchrotron radiation is self-absorbed with a turnover frequency of around 110~GHz. Opacity effects due to synchrotron self-absorption can result in frequency-dependent time lags with shorter time lags at higher frequencies \citep{Fuhrmann+2014}. 

We note negative time lags in the correlation between {radio and $\gamma$-ray light curves during P2}. Previous studies have reported that this phenomenon occurs due to the interaction of a traveling shock with a stationary shock in a relativistic jet \citep[e.g.][]{Agudo+2011, Ramakrishnan+2014}. We conducted a jet kinematics analysis to explore the possibility of the shock-shock interaction in Sect.~\ref{sec: jet_kinematics}. In summary, our analysis suggests the presence of radio counterparts for two $\gamma$-ray flares (Flare 3 and Flare 6).

We found 3$\sigma$ correlations between the $\gamma$-ray and optical light curves in P2 and P3. We observed zero time lags during P2 (Fig.~\ref{fig: DCF_P2}), which is consistent with the $\gamma$-optical correlations over the entire period. However, we found negative time lags of $\sim$10~days between the $\gamma$-ray and optical light curves during P3 (Fig.~\ref{fig: DCF_Flare6_radioOpt-gamma-ray}). This implies that the optical emission preceded the $\gamma$-ray emission. {A detailed discussion of the interpretation of negative time lags is presented in} Sect.~\ref{subsec: gamma-optical correlation}.

\section{Jet kinematics}     \label{sec: jet_kinematics}

\begin{figure}[htb!]
\includegraphics[scale=0.5]{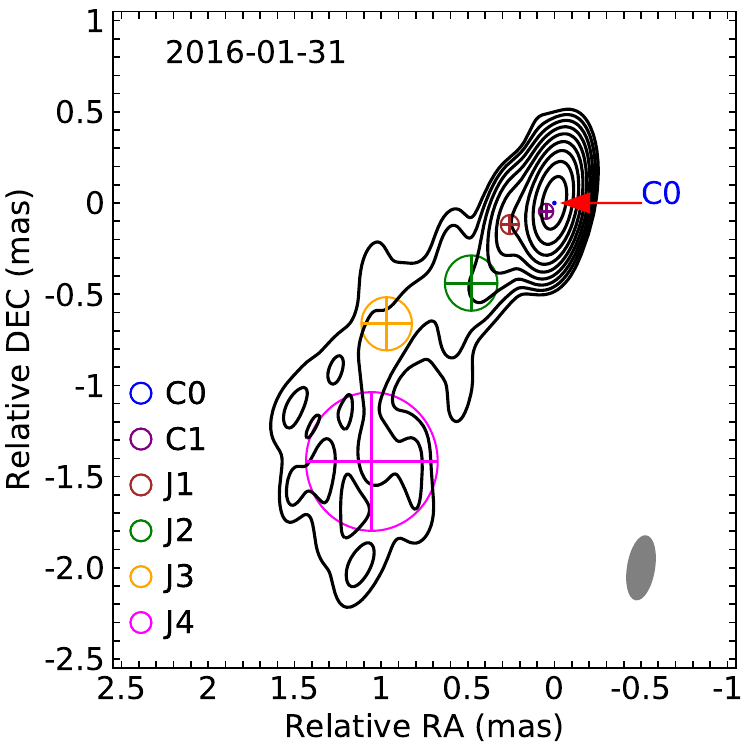}
\centering
\caption{A VLBA 43~GHz image of CTA~102 {taken} on 2016 January 31. The contours increase by a factor of two from the lowest level of 3.92~mJy~${\rm beam}^{-1}$. The map peak is 1.60~Jy~${\rm beam}^{-1}$. The grey ellipse in the bottom right corner shows the restoring beam of $0.35 \times 0.15$~mas at -10~deg. The circles with crosses {indicate} the position{s} and size{s} of the core (C0) and the jet components (C1 and J1--4). \label{fig: BU_map_2016}}
\end{figure}

For the analysis of jet kinematics, we followed the formalism of \citet{Jorstad+2017} and \citet{Weaver+2022}. To parameterize the compact bright features in the jet of CTA~102, we performed a model-fitting analysis of VLBA 43~GHz data taken from 2016 January 1 to 2018 August 26 (MJD 57388--58356), using the \emph{MODELFIT} task in \texttt{DIFMAP}. We fit circular two-dimensional Gaussians to CLEAN images, finding multiple jet components for each epoch. The fits also yield jet parameters for the components, including the flux density, the relative distance from the core ($R$), the size, and the relative position angle with respect to the core (PA). 
{We calculate uncertainties of jet parameters using formulas given by \citet{Jorstad+2017}.}
An example is provided in Fig.~\ref{fig: BU_map_2016}, which displays the fit of Gaussian components superimposed on the CLEAN image taken on 2016 January 31. The core (labelled C0) is the brightest region {presumed to be stationary} at the upper end of the jet {(i.e. the origin of each map)}. We labelled the jet components close to the core as C1--3 and the others as J1--4. The model fitting parameters of the core and jet components and their uncertainties are available on Zenodo\footnote{\url{https://doi.org/10.5281/zenodo.14642338}}.

   \begin{table}
      \caption[]{Motions of the moving jet components}
         \label{tab: motions_jet}
      \centering
      \tiny
        \begin{tabular}{lccccc}
            \hline
            \hline
            \noalign{\smallskip}
            Name & $t_{\rm start}$  &  $t_{\rm end}$ & $\mu$ & $\beta_{\rm app}$ & $t_{\rm 0}$      \\
            {}   & (MJD)            & (MJD)          & (mas/yr)    & (c)                     & (MJD) \\
            \noalign{\smallskip}
            \hline
            \noalign{\smallskip}
            C2   & 57720 & 58063 & $0.18 \pm 0.01$ & $9.8 \pm 0.6$ & $57545^{+24}_{-23}$ \\
            C3   & 58187 & 58356 & $0.32^{+0.04}_{-0.05}$ & $17.8^{+2.0}_{-2.7}$ & $58151^{+17}_{-16}$ \\
            \noalign{\smallskip}
            \hline
        \end{tabular}
        \tablefoot{The columns are (1) {identification} of the jet component, (2) the first detection time of the jet component in MJD, (3) the last detection time of the jet component in MJD, (4) proper motion in mas/yr, (5) {apparent speed} in {units of c}, and (6) time of ejection of the jet component in MJD.}
    \end{table}

\begin{figure}[htb!]
\includegraphics[scale=0.5]{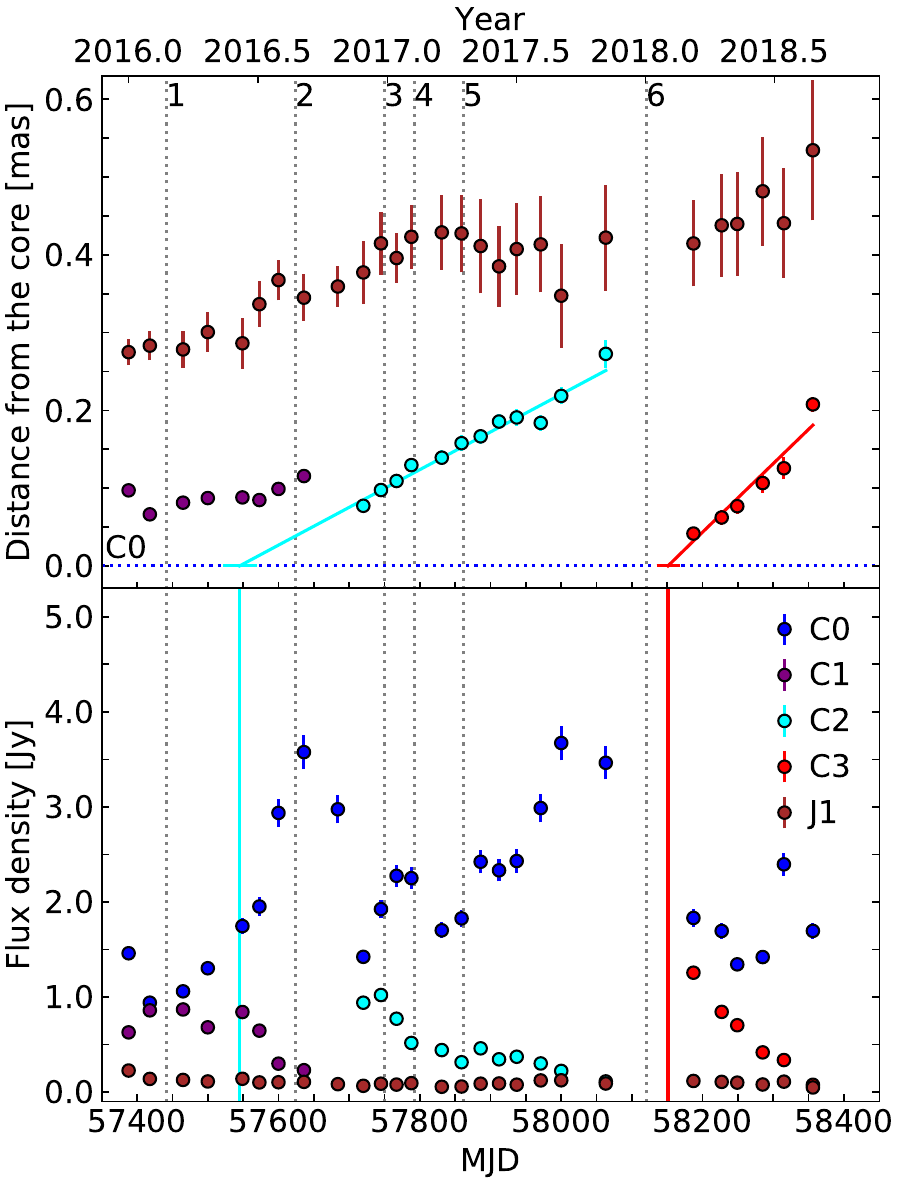}
\centering
\caption{Top: Separation{s} of the 43~GHz jet components from the core ({$R < 0.5$~mas}).
The grey vertical lines indicate {the peaks of the $\gamma$-ray flares}.
The {blue} dotted line shows the position of C0.
The cyan and red solid lines represent linear fits to the positions of C2 and C3, respectively.
The cyan and red horizontal lines correspond to the uncertainties in the times of ejection of C2 and C3, respectively.
Bottom: Light curves of the 43~GHz core and jet components.
The cyan and red vertical lines indicate the ejection times of C2 and C3, respectively.}
\label{fig: BU_distance+LC}
\end{figure}

We examined the jet components close to the core, at $R < 0.5$~mas. The top panel of Fig.~\ref{fig: BU_distance+LC} displays the separations of jet components from the core as a function of time. Over the course of the data period, {a jet component C1 showed a quasi-stationary at $\sim$0.1~mas from the core, which has been consistently reported in multiple studies \citep[e.g.][]{Jorstad+2005, Fromm+2013a, Casadio+2015, Jorstad+2017, Casadio+2019, Weaver+2022}. Additionally, we found that} two of the jet components, C2 and C3, moved farther downstream from the core. The bottom panel of Fig.~\ref{fig: BU_distance+LC} presents the light curves of the core and jet components of the source.
{To determine the motions and physical parameters for each jet component, we fitted the measurements of $R$ for components C2 and C3 with a linear function and obtained the best-fit solution of the proper motion $\mu$ in mas/yr.}
The apparent speed, $\beta_{\rm app}$, was obtained by using the relation that a proper motion of 1~mas/yr corresponds to 55.2c (see Sect.~\ref{sec: intro}). The ejection time of a jet component, $t_0$, is defined as the time when the centroid of the jet component coincides with the centroid of the core.
{We determined $t_0$ by extrapolating to $R=0$ from the fitted linear motion.} The uncertainty in $t_0$ was computed by propagating errors from the fit results. 
We added a minimum uncertainty of {0.005/$\mu$~yr} to the uncertainty in $t_0$, where 0.005~mas is the half-size of a pixel in the VLBA images \citep{Weaver+2022}. 
Table~\ref{tab: motions_jet} presents the motions of the moving jet components we calculated.

\begin{figure*}[htb!]
\includegraphics[scale=0.65]{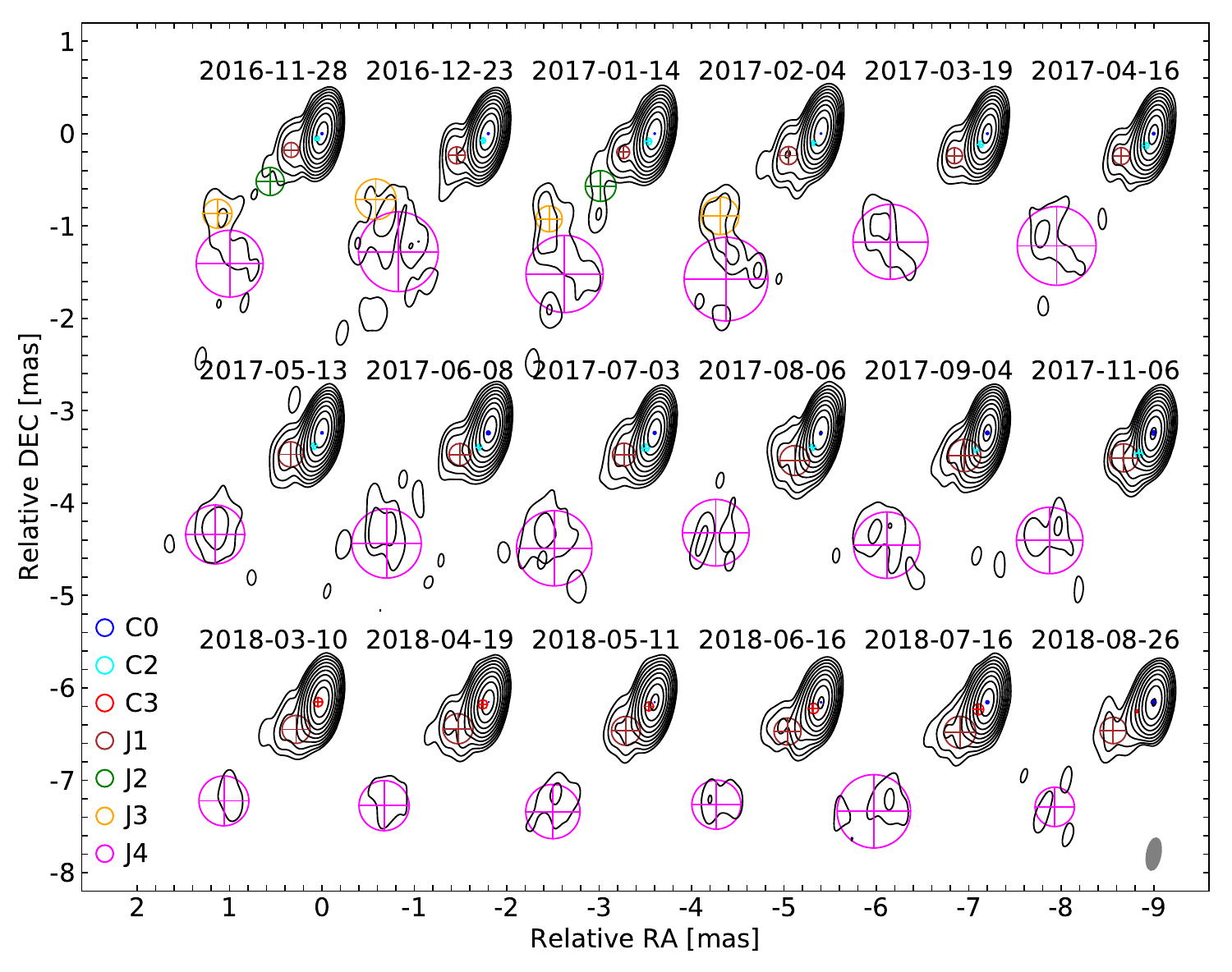}
\centering
\caption{VLBA 43~GHz images of CTA~102 in {2016--2018}. The epoch of the observations is presented above each contour map. The contours increase by a factor of two from the lowest level of 5.90~mJy~${\rm beam}^{-1}$. All maps are restored with a common beam that is the same as in Fig.~\ref{fig: BU_map_2016} (illustrated at the bottom right corner). 
The circles with crosses {indicate} the position{s} and size{s} of the core (C0) and the jet components {(C2, C3, J1, J2, J3, and J4)}. \label{fig: BU_maps_2016-2018}}
\end{figure*}

We derived the physical parameters of the moving jet components.
To estimate the variability {timescale} of $\tau_{\rm var}$, we {followed \citet{Jorstad+2017} in using} a model with an exponential decay in the form of {$\ln{S(t)} \propto -t$}.
{Radio} flares from blazar jets{,} including CTA~102{, have been shown to follow} exponential profiles \citep[e.g.][]{Teraesranta+1994, Rani+2013, Liodakis+2018a, KimSH+2022}.
{Individual} exponential flares are related to the emergence of new jet components {in} VLBI observations{,} and the decaying profile of the flux {density is} similar to that of the total flux density \citep{Valtaoja+1999, Savolainen+2002}.
We note that \citet{KimSH+2022} identified seven exponential flares in the 15~GHz light curve{,} and two of them are closely connected with the emergence of VLBI jet components whose decaying {flux density profile} resembles that of the 15~GHz flux density (see below).
Thus, the {decay in flux of a jet component} can be approximated as $\ln{(S(t)/S_o)} = k(t - t_{\rm max})$, where $S(t)$ is the flux density as a function of time, $t_{\rm max} (\geq t$) is the time corresponding to {the maximum flux density of $S_{\rm max}$}, $S_0$ is the flux density of the fit to the light curve decay at $t_{\rm max}$, and $k$ is the slope of the fit.
The variability {timescale} is computed to be $\tau_{\rm var} = |1/k|$.
To calculate the uncertainty in $\tau_{\rm var}$, we used {\emph{EMCEE}, a \texttt{PYTHON} package that} employs the Markov Chain Monte Carlo technique \citep{Foreman-Mackey+2013}.
The variability Doppler factor is given by
\begin{equation}	\label{eq: Doppler factor}
    {\delta_{\rm var} = \frac{K s D}{\tau_{\rm var} (1+z)},}
\end{equation}
where {$K = 15.8~{\rm yr~{mas}^{-1}~{Gpc}^{-1}}$}, $D$ is the luminosity distance in Gpc, $\tau_{\rm var}$ is in the unit of years, and $s$ is the angular size of the jet component in mas, which is 1.6 times the full width at half maximum of the Gaussian modelled jet component assuming that the actual shape of the jet component is similar to a uniform face-on disk. 
Using the apparent speed and the variability Doppler factor, the Lorentz factor ({$\Gamma_{\rm var}$}) and the viewing angle ({$\theta_{\rm var}$}) with respect to the line of sight {are given by}
\begin{equation}
    {\Gamma_{\rm var}} = \frac{\beta_{\rm app}^2 + \delta_{\rm var}^2 + 1}{2 \delta_{\rm var}}, \\
\end{equation}
\begin{equation}
    {\theta_{\rm var}} = \mathrm{atan}(\frac{2 \beta_{\rm app}}{\beta_{\rm app}^2 + \delta_{\rm var}^2 - 1}).
\end{equation}
Table~\ref{tab: physical_parameters_jet} presents the physical parameters of the jet components.

   \begin{table}
      \caption[]{Physical parameters of the moving jet components}
         \label{tab: physical_parameters_jet}
      \centering
        \begin{tabular}{lcccc}
            \hline
            \hline
            \noalign{\smallskip}
            Name & $\tau_{\rm var}$ & $\delta_{\rm var}$ & $\Gamma_{\rm var}$ & $\theta_{\rm var}$ \\
            {}   & ({yr})            & {}                 & {}                     & (deg) \\
            \noalign{\smallskip}
            \hline
            \noalign{\smallskip}
            C2   & $0.51^{+0.09}_{-0.08}$ & $12 \pm 2$ & $10.1 \pm 0.6$ & $4.5^{+1.0}_{-0.9}$ \\ 
            C3   & $0.22^{+0.03}_{-0.04}$ & $38 \pm 6$ & $23.1^{+2.5}_{-2.7}$ & $1.2 \pm 0.3$  \\
            \noalign{\smallskip}
            \hline
        \end{tabular}
        \tablefoot{The columns are (1) identification of the jet component, (2) e-folding rise time in years, (3) variability Doppler factor, (4) bulk Lorentz factor, and (5) viewing angle with respect to the line of sight in degrees.}
    \end{table}

Fig.~\ref{fig: BU_maps_2016-2018} displays the VLBA~43 GHz images of the source in 2016--2018, including detections of C2 and C3. C2 was first detected on 2016 November 28 (MJD~57720) at a distance of $0.077 \pm 0.007$~mas from the core. {It then travelled down the jet with an apparent speed of $\beta_{\rm app} = 9.8 \pm 0.6$.} The variability Doppler factor of C2 is {$\delta_{\rm var} = 12 \pm 2$}. C2 was ejected in a {50~day window around 2016 June 5 (${\rm MJD}~57545^{+24}_{-23}$)}. {Our estimate of $t_0$ is roughly consistent with that of \citet{Casadio+2019} within the uncertainty.} 

The jet component C3 was first detected on 2018 March 10 (MJD~58187) at a distance of $0.042 \pm 0.009$~mas from the core. {Emerging from the core, C3 moved downstream the jet with an apparent speed of $\beta_{\rm app} = 17.8^{+2.0}_{-2.7}$. The variability Doppler factor of C3 is $\delta_{\rm var} = 38 \pm 6$.} We note that this high estimate of the Doppler factor is consistent with those in previous studies {\citep[e.g.][]{Casadio+2015, Casadio+2019}}. C3 was ejected in {a 33~day window around 2018 February 1 (${\rm MJD} 58151^{+17}_{-16}$).}

\section{Discussion} \label{sec: discussion}

\subsection{\texorpdfstring{$\gamma$}{Lg}-ray flares in shock-shock interaction regions}    \label{subsbec: shock-shock interaction}

We discuss the origin of the $\gamma$-ray flares based on the jet kinematics. The motion of C2, presented in Sect.~\ref{sec: jet_kinematics}, shows that the passage of C2 through C1 (at $\sim$0.1~mas from the core) coincides with the peak of Flare~3 on 2016 December 27 (MJD~57750), as well as the peaks of optical and X-ray {local maxima} during P2. The peak flux density of C2 matches the peak of Flare~3, indicating that the high-energy {activity is} likely a result of the shock-shock interaction between C1 and C2 \citep[see also,][]{Casadio+2019}.
The $\gamma$-radio correlation analysis yielding negative time lags of $< -50$~days supports the scenario. During P2, the core flux density reached a peak of $3.7 \pm 0.2$~Jy on MJD~57636 (bottom in Fig.~\ref{fig: BU_distance+LC}), accompanied by the size variation with a factor of $\sim$5 increase (from $\sim$0.01 to $\sim$0.07~mas).
The increase in both the flux density and size of the core can be attributed to the ejection of C2 emerging from the core (see Fig.~\ref{fig: BU_distance+LC}). In addition, the variation in the core flux density matches well with the radio {local maxima} in P2 (Fig.~\ref{fig: light curves+FP+BB}). The observed negative time lags (the radio emission leading the $\gamma$-ray emission) confirm that Flare~3 occurred downstream of the radio core.

\begin{figure}[htb!]
\includegraphics[scale=0.59]{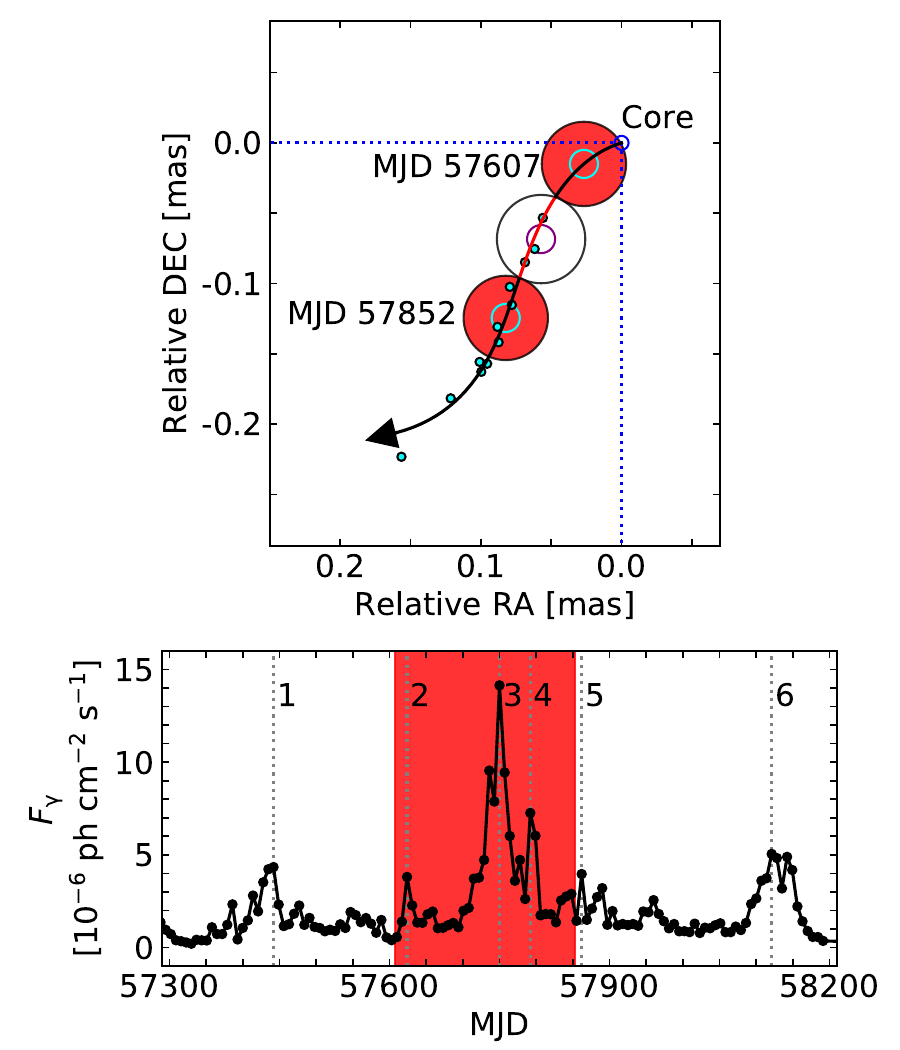}
\centering
\caption{{Top: Vector motion fit and sky position of C2. The position of C2 is relative to the core, denoted by the blue dotted lines. The black solid line represents the vector fit to the separation from the core for C2. The position at each epoch is denoted by filled cyan circles while the positions inferred from the motion of C2 are shown with unfilled large cyan circles. 
{The unfilled purple circle represents the mean separation of C1 from the core.}
The filled red circles indicate the size of C2 assuming its lower limit, during the period of the shock-shock interaction. The red part in the solid line indicates the interaction region between C1 and C2. Bottom: $\gamma$-ray light curve during the flaring period (MJD~57300--58200). The shaded red region represents the period of the interaction between C1 and C2.}}
\label{fig: Trajectory_C2+LC_Gamma}
\end{figure}

Furthermore, we examined the interaction between C1 and C2 by modelling the trajectory of C2 using its $R$ and PA. Fig.~\ref{fig: Trajectory_C2+LC_Gamma} presents the trajectory of C2 and the mean separation from the core for C1 (top) and the $\gamma$-ray light curve (bottom). We used the linear-fit result of C2 as presented in Sect.~\ref{sec: jet_kinematics} and fitted a second-order polynomial function to the PA of C2. For the size of C2, we took the minimum value. The analysis revealed that the traveling shock (C2), propagating along a helical trajectory (top in Fig.~\ref{fig: Trajectory_C2+LC_Gamma}), passed through C1 beginning at least MJD~57607 and exited from the region of influence on the shock-shock interaction at MJD~57852. The period of the shock-shock interaction is estimated to be around 250 days, coinciding with three $\gamma$-ray flares (Flares~2--4). This implies that the shock-shock interaction likely triggered not only Flare~3 but also other major $\gamma$-ray flares during P2. Throughout the interaction between C1 and C2, the separation from the core for C2 varied from $\sim$0.02 to $\sim$0.16~mas. Assuming a viewing angle of $\theta \sim 4.5$~deg taken from C2, the angular separation corresponds to the de-projected distance (from the core) of $\sim$2--17~pc. Therefore, we suggest the possibility that the $\gamma$-ray flares during P2 occurred parsecs downstream of the core. Our results are consistent with results estimated in previous studies of blazars \citep[e.g.][]{Leon-Tavares+2011, Schinzel+2012, Rani+2015, Kramarenko+2022}.

\begin{figure}[htb!]
\includegraphics[scale=0.59]{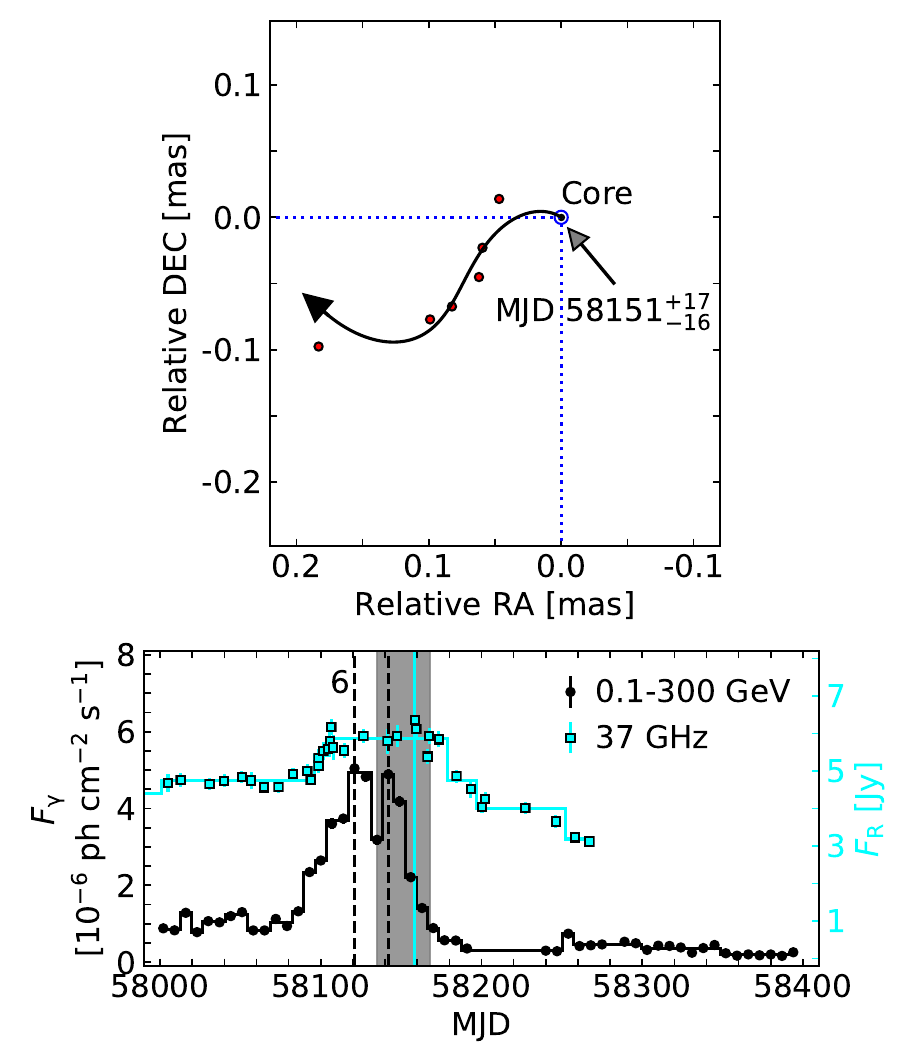}
\centering
\caption{Top: Same as Fig.~\ref{fig: Trajectory_C2+LC_Gamma} for the jet component C3. The grey arrow indicates the ejection position of C3. Bottom: $\gamma$-ray (black) and 37~GHz (cyan) light curves during P3 (MJD~58000--58400). {The black and cyan solid lines represent the Bayesian blocks of the $\gamma$-ray and 37~GHz light curves, respectively.} The black {vertical} dashed lines indicate peaks of {two $\gamma$-ray local maxima (including Flare~6) in the Bayesian block analysis}. The cyan vertical line denotes the peak of the 37~GHz flux density. The shaded grey region shows the ejection of C3.}
\label{fig: BU_distance+LC_FP4}
\end{figure}

Fig.~\ref{fig: BU_distance+LC_FP4} displays the trajectory of C3 (top) and the $\gamma$-ray and 37~GHz light curves with the Bayesian blocks {(bottom)} during P3. The Bayesian block analysis reveals that the $\gamma$-ray light curve during P3 is characterized by two {local maxima} at MJD~58121 (Flare~6) and MJD~58142. {The second local maximum} coincides with the ejection time of C3, estimated to be ${\rm MJD}~58151^{+17}_{-16}$ within the uncertainty. The lack of 43 GHz VLBA data near the ejection time of C3 makes it challenging to study the relationship between the variation in the core flux density and the ejection of C3 from the core. \citet{KimSH+2022} reported a strong correlation between the core flux variability and the single-dish flux variability of the jet in CTA 102. During P3, a local maximum at MJD~58158 in the 37~GHz light curve is consistent with the ejection time of C3. The time coincidence of the {local maxima} with the shock ejection suggests that the interaction between C3 and the core (possibly a standing shock) may have been associated with the $\gamma$-ray activity during P3. Similar to C2, C3 propagates down the jet along a helical trajectory due to the large variation in PA.

In order to constrain the location of the $\gamma$-ray flaring region, we use the multi-wavelength time lags between the radio and $\gamma$-ray light curves. The distance between the $\gamma$-ray emission region and the radio core can be calculated as $\Delta r_{\rm radio-\gamma} = \beta_{\rm app} c t^{\rm sour}_{\rm radio, \gamma} / {\sin{\theta}}$, where $t^{\rm sour}_{\rm radio, \gamma}$ is the radio to $\gamma$-ray time delay in the source frame {\citep[e.g.][]{Pushkarev+2010, Ramakrishnan+2014, KimDW+2022, Cheong+2024}}. We computed this time delay following the relation of $t^{\rm sour}_{\rm radio, \gamma} = t^{\rm obs}_{\rm radio, \gamma} (1+z)^{-1}$, where $t^{\rm obs}_{\rm radio, \gamma}$ is the radio to $\gamma$-ray time delay in the observer frame. Regarding the association of the ejection epoch for C3 with the peak of the $\gamma$-ray flare, we used the physical jet parameters (i.e. $\mu$, $\beta_{\rm app}$, and {$\theta_{\rm var}$}) of C3.
{We found that the $\gamma$-ray flaring regions are located at $\Delta r_{\rm \gamma,15GHz} = 8.1 \pm 2.8$~pc and $\Delta r_{\rm \gamma,37GHz} = 4.6 \pm 2.2$~pc upstream of the radio core.}

{The analysis of the frequency-dependent shift of the VLBI core location by \cite{Fromm+2015} indicates that the de-projected separation between the SMBH and the 86~GHz core is $7.1 \pm 3.3$~pc. Using the results of the core shift analysis from \cite{Fromm+2015}, the core shift values relative to 86~GHz are $\Delta r_{\rm 86,15~GHz} = 28.3 \pm 12.5$~pc, $\Delta r_{\rm 86,37~GHz} = 7.2 \pm 6.8$~pc, and $\Delta r_{\rm 86,43~GHz} = 5.2 \pm 6.2$~pc. The location of the radio core (from the SMBH) can be estimated to be $35.4 \pm 12.9$~pc, $14.3 \pm 7.6$~pc, and $12.3 \pm 7.0$~pc at 15, 37, and 43~GHz, respectively. These distances correspond to $\sim$$10^{4-5}$ gravitational radii, which are comparable to those estimated for other blazars \citep[e.g.][]{Marscher+2008, Abdo+2010a}.}

We discuss the seed photons for the physical mechanism that contributes to the $\gamma$-ray flares of CTA~102. {Given the location of the 43~GHz core relative to the central region and the traveling shock (C2) from the core}, the shock-shock interaction region associated with the $\gamma$-ray flares in 2016--2017 is located at {tens of parsecs} from the central engine, which is well beyond the BLR \citep[$\ll 1$~pc,][]{Pian+2005} {and the dust torus \citep[$\sim$2~pc,][]{Nenkova+2008}. This suggests that the sources of seed photons to generate $\gamma$-rays may come from the jet itself.}

{The 2018 $\gamma$-ray flaring region from the SMBH is estimated to be $9.7 \pm 7.9$~pc. This location is far from the BLR, but it could be comparable to the size of the torus, suggesting that the dust torus may have provided the seed photons for the EC mechanism during Flare~6.} The core position could vary due to flare events and newly ejected jet components propagating along the jet \citep[e.g.][]{Plavin+2019}. The typical variability of the core position has been reported to be $\sim$2~pc \citep{Plavin+2019}, which is not likely to have significant effects on our results.

\subsection{\texorpdfstring{$\gamma$}{Lg}-ray/optical correlations} \label{subsec: gamma-optical correlation}

We investigated the relationship between the $\gamma$-ray and optical fluxes to understand the emission mechanisms of the $\gamma$-ray flares. Most cases of correlations between $\gamma$-ray and optical {light curves} were found to be significant with confidence levels $> 99.7$\% (Sect.~\ref{subsec: DCF_results}). In a simple leptonic scenario (with constant magnetic field strength and viewing angle for a flaring period), it is assumed that synchrotron radiation from the jet is mostly emitted at optical wavelengths, whereas the $\gamma$-ray flaring emission is from IC scattering of optical/IR photons by relativistic electrons in the jet \citep[see, e.g.][]{Larionov+2016, Liodakis+2019, Sahakyan+2022}. Under this assumption, the relation between the flux density of the IC scattered emission ($F_{\rm IC}$) and that of the synchrotron radiation ($F_{\rm sync}$) is $F_{\rm IC} \propto F_{\rm sync}^m$, where $m$ is a slope of the flux--flux relation. If the dominant $\gamma$-ray emission mechanism is EC, then the only component of the synchrotron emission that also contributes to the IC emission is the population of relativistic electrons, in which case, $m = 1$. If it is SSC, both the relativistic electrons and their associated synchrotron photons contribute to the IC emission, resulting in an additional power (i.e. $m=2$) in the flux--flux relation.

\begin{figure}[htb!]
\includegraphics[scale=0.4]{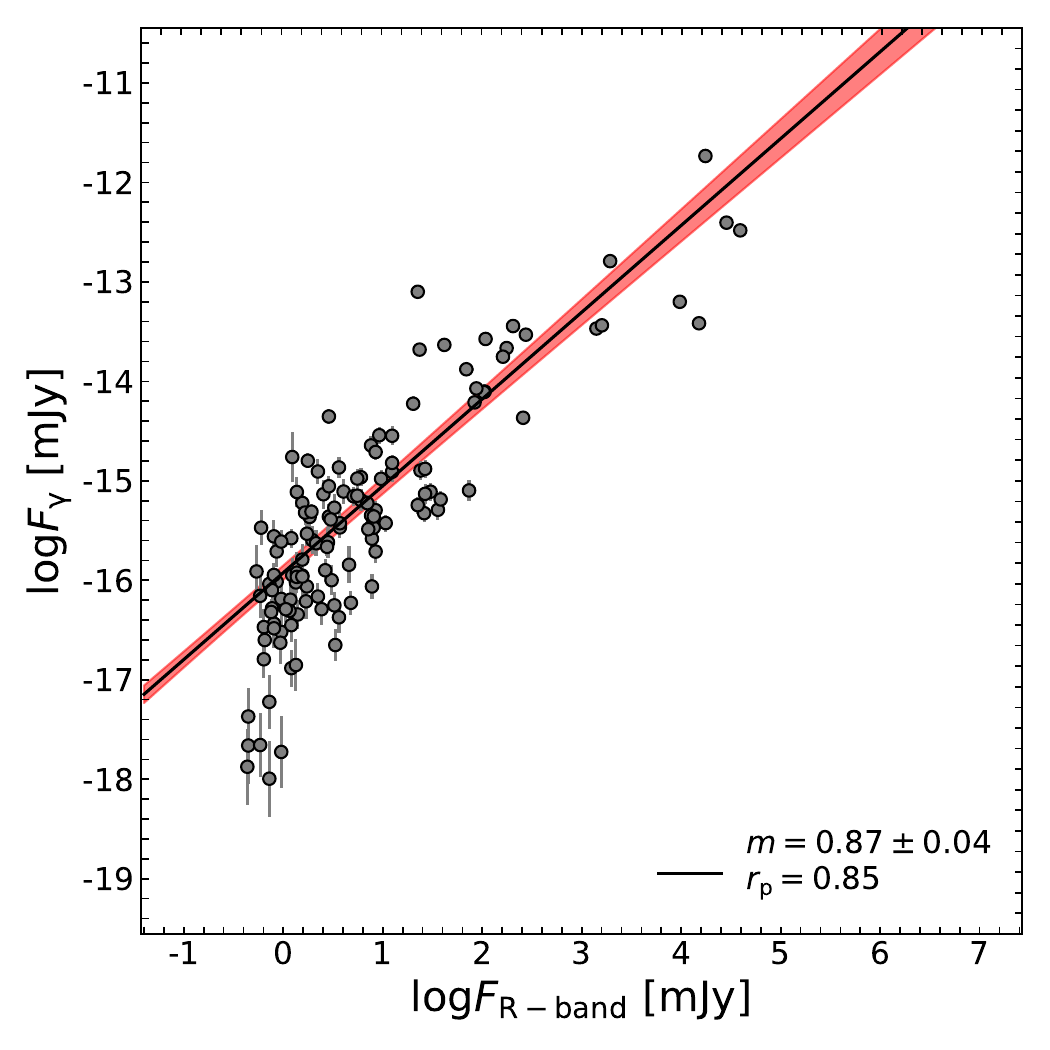}
\centering
\caption{The quasi-simultaneous $\gamma$-ray vs. optical (\textsl{R}-band) fluxes in log-log space. 
The data was taken from the six-year observation period in $\gamma$-rays (MJD~56200--58400).
{The best-fit solution of the flux--flux relation shows $F_{\rm IC} \propto F_{\rm sync}^{0.87 \pm 0.04}$ (plotted in black and red).}}
\label{fig: flux-correlation_gamma-optical_entire}
\end{figure}

\citet{D'Ammando+2019} investigated the relationship between the gamma-ray and optical fluxes on a logarithmic space during MJD~56293--57793 (including P1--2), yielding $F_{\rm IC} \propto F_{\rm sync}^{0.82}$ with a Pearson correlation coefficient of $r_{\rm p} = 0.95$. We conducted the analysis during the entire period (MJD~56200--58400). For this, we converted the unit of the $\gamma$-ray flux density (${\rm ph} \cdot {\rm cm}^{-2} \cdot {\rm s}^{-1}$) into mJy to be matched with that of the optical flux density. The optical flux density was shifted close in time with respect to the $\gamma$-ray flux density, regarding $\tau_{\rm DCF} = 3 \pm 5$~days. As Seen in Fig.~\ref {fig: flux-correlation_gamma-optical_entire}, we found a slope of $m = 0.87 \pm 0.04$ with $ r_{\rm p} = 0.85$, which is consistent with the result of \citet{D'Ammando+2019}. The slope close to the unity corresponds to the EC mechanism under the simple scenario.

{Considering the $\gamma$-ray emission site near the dust torus during Flare~6 (Sect.~\ref{subsbec: shock-shock interaction}), we assume that a $\gamma$-ray photon is produced by the upscattering of an infrared (IR) photon by a relativistic electron, expressed as $\epsilon_{\rm s} \gtrsim \gamma^2_{e} \epsilon_{\rm i}$ \citep{Boettcher+2012}, where $\epsilon_{\rm s}$ is the scattered $\gamma$-ray photon energy, $\gamma_{e}$ is the electron Lorentz factor, and $\epsilon_{\rm i}$ is the incident IR photon energy. \cite{Geng+2022} estimated the maximum $\gamma$-ray photon energy during Flare~6 to be $\epsilon_{\rm s} = 30.9$~GeV. The IR photon energy is predicted to be $\sim$$10^{2-3}$~eV, corresponding to a wavelength of 1.7--78.6~$\mu {\rm m}$ \citep{Malmrose+2011}. This suggests $\gamma_{e} \lesssim 10^{4}$, which is roughly consistent with the predictions from the SED analysis of blazars \citep{Kang+2014, Pandey+2024}.}

However, we note that a unity slope is possible within the SSC framework if variations in the magnetic field or the Doppler factor (resulting from changes in the jet viewing angle) are considered \citep{Larionov+2016, Liodakis+2019, D'Ammando+2019, Acciari+2020}. {The multi-wavelength light curve analysis yielded} a viewing angle of $\sim$8~deg in the 15~GHz band and $\sim$2~deg in the optical or $\gamma$-ray bands \citep{D'Ammando+2019}, suggesting that the unity slope for the $\gamma$-ray flares in P1--2 (2016--2017) may be due to changes in the Doppler factor resulting from variations in the jet viewing angle.
{The $\gamma$-ray variability during Flare~6 may be associated with these variations. We found a variation in the viewing angle from $\sim$4.5~deg to $\sim$1.2~deg (Table~\ref{tab: physical_parameters_jet}), as well as the large variation in the PA of C3 from $89.1 \pm 5.0$~deg to $129.2 \pm 4.1$~deg. Assuming that the observed flux density ($F_{\nu}$) is amplified due to the Doppler boosting effect, $F_{\nu}$ is related to the flux density in the jet frame ($F^{'}_{\nu}$) by $F_{\nu} = \delta^{3-\alpha} F^{'}_{\nu}$, where $F_{\nu} \propto \nu^{\alpha}$, $\delta$ is the Doppler factor defined as $\delta = 1/\Gamma(1-\beta \cos{\theta})$, $\alpha$ is the optically thin spectral index ($\alpha < 0$), $\beta$ is the intrinsic velocity of the jet, and $\Gamma$ is the bulk Lorentz factor defined as $\Gamma = 1/\sqrt{1 - \beta^2}$ \citep{Blandford+1979}. The ratio between the maximum and minimum flux densities in the optical during Flare~6 is $\sim$12. We thus obtain a ratio between the maximum and minimum Doppler factors of $\delta_{\rm max}/\delta_{\rm min} < 2.3$. The maximum Doppler factor of C3 is $\delta_{\max} \sim 44$ at $\theta \sim 0$~deg due to $\Gamma_{\rm var} \approx 23$ (see Table~\ref{tab: physical_parameters_jet}). This yields a lower limit of $> 19$ for the minimum Doppler factor at $\theta < 3$~deg.}


\begin{figure}[htb!]
\centering
\includegraphics[scale=0.5]{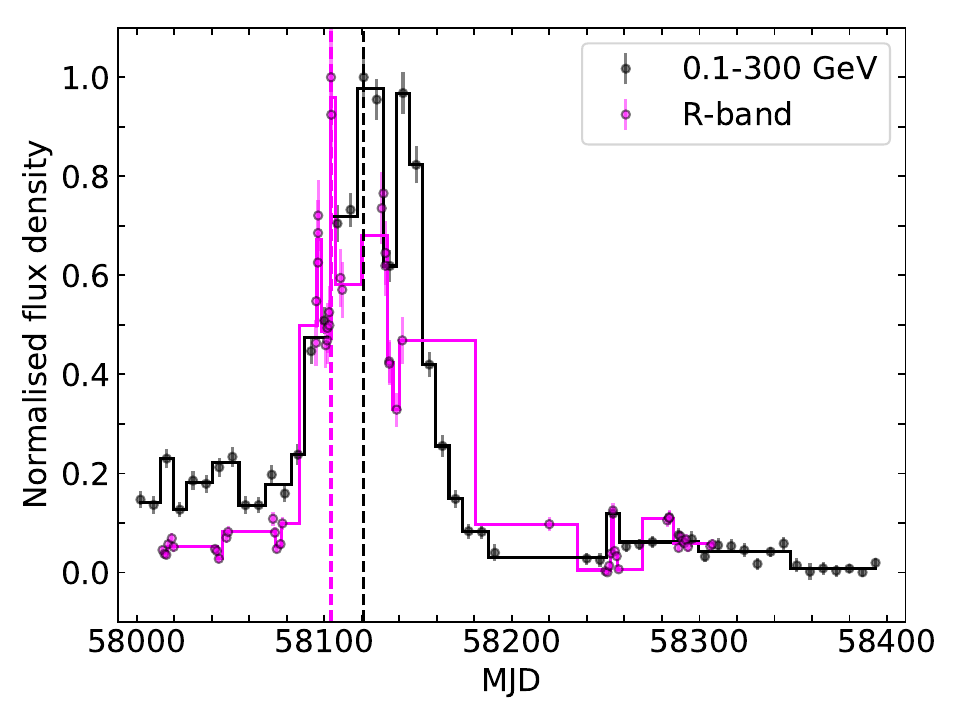}
\caption{{The $\gamma$-ray and optical (\textsl{R}-band) light curves during P3. The flux densities have been normalised to the maximum value in each band. The solid lines correspond to the Bayesian block representation of the light curves. The vertical dashed lines indicate peaks of the light curves.}}
\label{fig: light curves+BB_gamma-opt_P3}
\end{figure}

We discuss the observed negative time lags between the $\gamma$-ray and optical light curves during P3 (Sect.~\ref{subsec: DCF_results}), {which have been reported in previous studies as well \citep[e.g.][]{Abdo+2010c, Agudo+2011b, Hayashida+2012, Gupta+2017, Liodakis+2019}.} Fig.~\ref{fig: light curves+BB_gamma-opt_P3} displays the $\gamma$-ray and optical (\textsl{R}-band) light curves with normalised flux densities and the Bayesian block representation during P3. In this period, we identified seven optical and five $\gamma$-ray local maxima (including Flare~6).
{After applying a time shift equal to the observed time lag,}
the optical local maxima at MJD~58097 and MJD~58142 appear to have no $\gamma$-ray counterparts. The negative time lags are likely due to the cross-match of two optical local maxima at MJD~58104 and MJD~58132 with two $\gamma$-ray local maxima at MJD~58121 (Flare~6) and MJD~58142.
Assuming a co-spatial origin of optical and $\gamma$ emission, the optical local maxima at MJD~58132 and MJD~58142 seem to overlap with the two $\gamma$-ray local maxima.
{The presence of optical local maxima without $\gamma$-ray counterparts may be attributed to magnetic field changes leading to synchrotron emission modulation with steady IC emission \citep[e.g.][]{Chatterjee+2013, Cohen+2014}.}
High-cadence observations and modeling of the broadband SED {are needed} for a clear understanding of the $\gamma$-optical correlation.

{The observed negative time lags could be explained in the framework of the one-zone leptonic EC scenario.}
In this scenario, we assume that the same electron population produces the optical and $\gamma$-ray emission by the synchrotron and EC processes in the fast-cooling regime, that is, the electron cooling timescale is much shorter than the source light crossing timescale \citep[e.g.][]{Derishev+2007, Hayashida+2012, Janiak+2012}. The synchrotron and EC losses depend on the energy densities of the magnetic field and the external radiation field (e.g. the BLR, dusty torus, etc.), respectively. The energy densities could have different spatial distributions along the jet \citep[refer to Fig.~6 in][]{Janiak+2012}. If the magnetic field energy density decreases more rapidly than the external radiation field energy density, the EC ($\gamma$-rays) peak could be delayed relative to the synchrotron (optical) peak, resulting in a negative time lag.

\section{Summary}    \label{sec: summary}

The blazar CTA~102 exhibited intense $\gamma$-ray flaring activity from 2016 to 2018. We investigated the physical origins of $\gamma$-ray flares in CTA~102 by performing a cross-correlation analysis between its $\gamma$-ray light curve and lower-energy light curves from multi-wavelength data. Using VLBA observations, we explored the pc-scale jet activity during the flaring periods. We summarise our conclusions as follows: 
\begin{enumerate}
    \item Based on our criteria, we identified six {flares} in the $\gamma$-ray light curve. Assuming a power-law spectrum in the $\gamma$-ray band, we found changes in the spectral index, exhibiting a harder-when-brighter trend, during the flaring period from 2016 to 2018.
    
    \item We conducted a cross-correlation analysis between the $\gamma$-ray light curve and those in lower-energy bands for both the entire observation period {and the flaring period}. The analysis revealed that the time lags between the $\gamma$-ray light curve and the optical and X-ray light curves were nearly zero, implying a time coincidence and thus a co-spatial origin.

    \item The correlations between the $\gamma$-ray and radio light curves were more complex over long periods, possibly due to more rapid variability in the $\gamma$-ray band compared to the radio band. We divided the flaring period into three parts (P1--3) and explored the correlations between the $\gamma$-ray and radio light curves. There are no significant correlations between the $\gamma$-ray and radio light curves in P1. The brightest $\gamma$-ray flare (Flare~3) showed significant correlations with the radio (15 and 37~GHz) local maxima as well as a negative time lag, indicating that the $\gamma$-ray emission lagged behind the radio emission. The most recent $\gamma$-ray flare (Flare~6) in P3 exhibited significant correlations with radio flares, with positive time lags between the $\gamma$-ray and radio light curves. This implies that the $\gamma$-ray emission preceded the radio emission.

    \item We used VLBA maps and found that two jet components (C2 and C3) were newly ejected from the core during P1 and P3, respectively. {Right after the ejection of C2}, there was an increase in both the flux density and size of the core. The variation in core flux density seems to be linked to {the radio local maxima} correlated with Flare~3. The interaction between the moving jet component (C2) and a quasi-stationary jet component C1 (at $\sim$0.1~mas) is associated with Flare~3. Negative time lags in the $\gamma$-radio correlation support the shock-shock interaction scenario. In addition to Flare~3, multiple $\gamma$-ray flares in P2 could be expected to occur as the traveling shock passes through the stationary one, at $\sim$2--17~pc downstream of the radio core.

    \item The ejection time of the jet component C3 coincided with the $\gamma$-ray activity in P3, suggesting a connection between the two phenomena. The $\gamma$-ray flaring region is located parsecs upstream of the radio core as inferred from multi-wavelength time lags between the radio and $\gamma$-ray light curves and physical jet parameters of C3. The $\gamma$-ray flaring regions are well beyond the BLR, which suggests that the sources of seed photons for $\gamma$-rays may come from {either the jet itself or the dust torus}.

    \item We found a linear relationship between the $\gamma$-ray and the optical fluxes, suggesting that the EC mechanism is at play. Alternatively, this relationship may be consistent with the SSC mechanism as well, provided that the jet viewing angle varies as the jet propagates.
\end{enumerate}

\section*{Data availability}
Appendix~A, containing the table used in this publication, is available on Zenodo at: \href{https://doi.org/10.5281/zenodo.14642338}{https://doi.org/10.5281/zenodo.14642338}.

\begin{acknowledgements}
{We thank the anonymous referee for useful comments that helped improve this manuscript.}
This research has made use of data from the OVRO 40-m monitoring program (Richards, J. L. et al. 2011, ApJS, 194, 29), supported by private funding from the California Institute of Technology and the Max Planck Institute for Radio Astronomy, and by NASA grants NNX08AW31G, NNX11A043G, and NNX14AQ89G and NSF grants AST-0808050 and AST- 1109911.      
This publication makes use of data obtained at the Mets\"ahovi Radio Observatory, operated by Aalto University in Finland.  
The IAA-CSIC group acknowledges financial support from the grant CEX2021-001131-S funded by MCIN/AEI/10.13039/501100011033 to the ``Instituto de Astrof\'isica de Andaluc\'ia-CSIC". Acquisition and reduction of the POLAMI data was supported in part by MICIN through grants PID2019-107847RB-C44 abd PID2022-139117NB-C44. The POLAMI observations were carried out at the IRAM 30m Telescope. IRAM is supported by INSU/CNRS (France), MPG (Germany) and IGN (Spain).  
{CC acknowledges support by the European Research Council (ERC) under the HORIZON ERC Grants 2021 programme under grant agreement No. 101040021.}     
The Submillimeter Array is a joint project between the Smithsonian Astrophysical Observatory and the Academia Sinica Institute of Astronomy and Astrophysics and is funded by the Smithsonian Institution and the Academia Sinica. We recognize that Maunakea is a culturally important site for the indigenous Hawaiian people; we are privileged to study the cosmos from its summit.     
This paper makes use of the following ALMA data: ADS/JAO.ALMA\#2011.0.00001.CAL. ALMA is a partnership of ESO (representing its member states), NSF (USA) and NINS (Japan), together with NRC (Canada), MOST and ASIAA (Taiwan), and KASI (Republic of Korea), in cooperation with the Republic of Chile. The Joint ALMA Observatory is operated by ESO, AUI/NRAO and NAOJ.  
This study makes use of VLBA data from the VLBA-BU Blazar Monitoring Program (BEAM-ME and VLBA-BU-BLAZAR; \url{http://www.bu.edu/blazars/BEAM-ME.html}), funded by NASA through the \textsl{Fermi} Guest Investigator Program. The VLBA is an instrument of the National Radio Astronomy Observatory. The National Radio Astronomy Observatory is a facility of the National Science Foundation operated by Associated Universities, Inc.  
Data from the Steward Observatory spectropolarimetric monitoring project were used. This program is supported by Fermi Guest Investigator grants NNX08AW56G, NNX09AU10G, NNX12AO93G, and NNX15AU81G.    
The \textsl{Fermi} LAT Collaboration acknowledges generous ongoing support
from a number of agencies and institutes that have supported both the
development and the operation of the LAT as well as scientific data analysis.
These include the National Aeronautics and Space Administration and the
Department of Energy in the United States, the Commissariat \`a l'Energie Atomique
and the Centre National de la Recherche Scientifique / Institut National de Physique
Nucl\'eaire et de Physique des Particules in France, the Agenzia Spaziale Italiana
and the Istituto Nazionale di Fisica Nucleare in Italy, the Ministry of Education,
Culture, Sports, Science and Technology (MEXT), High Energy Accelerator Research
Organization (KEK) and Japan Aerospace Exploration Agency (JAXA) in Japan, and
the K.~A.~Wallenberg Foundation, the Swedish Research Council and the
Swedish National Space Board in Sweden.
Additional support for science analysis during the operations phase is gratefully
acknowledged from the Istituto Nazionale di Astrofisica in Italy and the Centre
National d'\'Etudes Spatiales in France. This work performed in part under DOE
Contract DE-AC02-76SF00515.     
This work was supported by the National Research Foundation of Korea (NRF) grant funded by the Korea government (MIST) (2020R1A2C2009003).
{This work was supported by the National Research Foundation of Korea (NRF) grant funded by the Korea government (MSIT; RS-2024-00449206).
This research has been supported by the POSCO Science Fellowship of POSCO TJ Park Foundation.}
\end{acknowledgements}

\bibliographystyle{aa}
\bibliography{refs}

\end{document}


\title{Appendix: Gamma-ray flares from the jet of the blazar CTA 102 in 2016--2018}


   \titlerunning{Gamma-ray Flares from the Jet of CTA 102}
   \authorrunning{S. Kim et al. 2024}

   \author{Sanghyun Kim\inst{1,2}
   \orcidicon{0000-0001-7556-8504}
          \and
          Sang-Sung Lee\inst{1,2}\thanks{Corresponding author; \href{mailto:sslee@kasi.re.kr}{sslee@kasi.re.kr}}
          \orcidicon{0000-0002-6269-594X}
          \and
          Juan Carlos Algaba\inst{3}
          \orcidicon{0000-0001-6993-1696}
          \and
          Bindu Rani\inst{4,5}
          \orcidicon{0000-0001-5711-084X}
          \and
          Jongho Park\inst{6,7}
          \orcidicon{0000-0001-6558-9053}
          \and 
          Hyeon-Woo Jeong\inst{1,2}
          \orcidicon{0009-0005-7629-8450}
          \and 
          Whee Yeon Cheong\inst{1,2}
          \orcidicon{0009-0002-1871-5824}
          \and 
          Filippo D’Ammando\inst{8}
          \orcidicon{0000-0001-7618-7527}
          \and
          Anne L\"ahteenm\"aki\inst{9,10}
          \orcidicon{0000-0002-0393-0647}
          \and 
          Merja Tornikoski\inst{9}
          \orcidicon{0000-0003-1249-6026}
          \and 
          Joni Tammi\inst{9}
          \orcidicon{0000-0002-9164-2695}
          \and 
          Venkatessh Ramakrishnan\inst{9,11}
          \orcidicon{0000-0002-9248-086X}
          \and
          Iv\'an Agudo\inst{12}
          \orcidicon{0000-0002-3777-6182}
          \and
          Carolina Casadio\inst{13,14}      
          \orcidicon{0000-0003-1117-2863}
          \and
          Juan Escudero\inst{12}
          \orcidicon{0000-0002-4131-655X}
          \and
          Antonio Fuentes\inst{12}
          \orcidicon{0000-0002-8773-4933}
          \and
          Efthalia Traianou\inst{12}
          \orcidicon{0000-0002-1209-6500}
          \and
          Ioannis Myserlis\inst{15}
          \orcidicon{0000-0003-3025-9497}
          \and
          Clemens Thum\inst{15}
          }

    \institute{Korea Astronomy and Space Science Institute, 776 Daedeok-daero, Yuseong-gu, Daejeon 34055, Republic of Korea
         \and
             Department of Astronomy and Space Science, University of Science and Technology, 217 Gajeong-ro, Yuseong-gu, Daejeon 34113, Republic of Korea
         \and
             Department of Physics, Faculty of Science, {Universiti Malaya}, 50603 Kuala Lumpur, Malaysia
         \and
             NASA Goddard Space Flight Center, Greenbelt, MD 20771, USA
         \and
             Department of Physics, American University, Washington, DC 20016, USA
         \and
             School of Space Research, Kyung Hee University, 1732, Deogyeong-daero, Giheung-gu, Yongin-si, Gyeonggi-do 17104, Republic of Korea
         \and
             Institute of Astronomy and Astrophysics, Academia Sinica, P.O. Box 23-141, Taipei 10617, Taiwan, R.O.C.
         \and
             INAF – Istituto di Radioastronomia, Via Gobetti 101, I-40129 Bologna, Italy
         \and
             Aalto University Mets\"ahovi Radio Observatory, Mets\"ahovintie 114, 02540 Kylm\"al\"a, Finland
         \and
             Aalto University Department of Electronics and Nanoengineering, P.O. BOX 15500, FI-00076 AALTO, Finland
         \and
             Finnish Centre for Astronomy with ESO (FINCA), University of Turku, Vesilinnantie 5, 20014 University of Turku, Finland
         \and
             Instituto de Astrof\'isica de Andaluc\'ia-CSIC, Glorieta de la Astronom\'ia, E-18008, Granada, Spain
         \and
             Institute of Astrophysics, Foundation for Research and Technology - Hellas, Voutes, 7110 Heraklion, Greece
         \and    
             Department of Physics, University of Crete, 70013, Heraklion, Greece
         \and
             Institut de Radiostronomie Milim\'etrique, Avenida Divina Pastora, 7, Local 20, E-18012 Granada, Spain
             }      
   \date{Accepted January 13, 2025}


  \abstract
  {CTA 102 is a $\gamma$-ray bright blazar that exhibited multiple flares in observations by the Large Area Telescope on board the \textsl{Fermi Gamma-Ray Space Telescope} during the period of 2016--2018. We present results from the analysis of multi-wavelength light curves aiming at revealing the nature of $\gamma$-ray flares from the relativistic jet in the blazar. We analyse radio, optical, X-ray, and $\gamma$-ray data obtained in a period from 2012 September 29 to 2018 October 8. We identify {six} flares in the $\gamma$-ray light curve, showing a harder-when-brighter-trend in the $\gamma$-ray spectra. We perform a cross-correlation analysis of the multi-wavelength light curves. We find nearly zero time lags between the $\gamma$-ray and optical and X-ray light curves, implying a common spatial origin for the emission in these bands. We find significant correlations between the $\gamma$-ray and radio light curves as well as negative/positive time lags with the $\gamma$-ray emission lagging/leading the radio during different flaring periods. The time lags between $\gamma$-ray and radio emission propose the presence of multiple $\gamma$-ray emission sites in the source. As seen in 43~GHz images from the Very Long Baseline Array, two moving disturbances (or shocks) were newly ejected from the radio core. The $\gamma$-ray flares from 2016 to 2017 are temporally coincident with the interaction between a traveling shock and a quasi-stationary one at $\sim$0.1~mas from the core. The other shock is found to emerge from the core nearly simultaneous with the $\gamma$-ray flare in 2018. Our results suggest that the $\gamma$-ray flares originated from shock-shock interactions.}

   \keywords{Radiation mechanisms: non-thermal -- Gamma rays: galaxies -- Galaxies: active -- Galaxies: jets -- quasars: individual: CTA 102}

   \maketitle
%

\begin{appendix} 
\onecolumn
\section{Model fitting parameters}  \label{sec: appendix_mfit}
In Table~\ref{tab: vlba_mfit}, we present the fit parameters of the Gaussian modelled jet components obtained from VLBI imaging at 43~GHz.

\begin{longtable}{ccccccc}
\caption{Fit model parameters from 43~GHz data from the VLBA-BU-BLAZAR monitoring program  \label{tab: vlba_mfit}} \\
\hline
\hline
\noalign{\smallskip}
Date        & MJD   & ID   & Flux Density & Distance & Size   & PA    \\
yyyy mon dd & (day) & Name & (Jy)         & (mas)    & (mas)  & (deg) \\
\noalign{\smallskip}
\hline
\noalign{\smallskip}
\endfirsthead   
\noalign{\smallskip}
\caption{Continued.} \\
\hline
\hline
\noalign{\smallskip}
Date        & MJD   & ID   & Flux Density & Distance & Size   & PA    \\
yyyy mon dd & (day) & Name & (Jy)         & (mas)    & (mas)  & (deg) \\
\noalign{\smallskip}
\hline
\noalign{\smallskip}
\endhead    
\noalign{\smallskip}
\hline
\noalign{\smallskip}
\endfoot    
\noalign{\smallskip}
\hline
\noalign{\smallskip}
\endlastfoot    
\noalign{\smallskip}
2016 Jan 01      &       57388   &       C0      &       1.46$\pm$0.073  &       0.0$\pm$0.0     &       0.034$\pm$0.007 &       0.0$\pm$0.0     \\
        &               &       C1      &       0.628$\pm$0.032 &       0.097$\pm$0.009 &       0.073$\pm$0.012 &       142.6$\pm$2.1   \\
        &               &       J2      &       0.034$\pm$0.011 &       0.59$\pm$0.04   &       0.115$\pm$0.031 &       130.2$\pm$3.4   \\
        &               &       J3      &       0.07$\pm$0.013  &       1.122$\pm$0.088 &       0.337$\pm$0.044 &       124.3$\pm$4.2   \\
        &               &       J4      &       0.139$\pm$0.015 &       1.903$\pm$0.16  &       0.79$\pm$0.057  &       142.8$\pm$4.6   \\
2016 Jan 31     &       57418   &       C0      &       0.941$\pm$0.047 &       0.0$\pm$0.0     &       0.014$\pm$0.005 &       0.0$\pm$0.0     \\
        &               &       C1      &       0.86$\pm$0.044  &       0.066$\pm$0.008 &       0.083$\pm$0.012 &       133.5$\pm$3.0   \\
        &               &       J1      &       0.139$\pm$0.011 &       0.283$\pm$0.019 &       0.106$\pm$0.021 &       114.7$\pm$2.7   \\
        &               &       J2      &       0.055$\pm$0.013 &       0.651$\pm$0.09  &       0.304$\pm$0.045 &       132.4$\pm$7.4   \\
        &               &       J3      &       0.038$\pm$0.013 &       1.173$\pm$0.108 &       0.293$\pm$0.048 &       124.4$\pm$4.9   \\
        &               &       J4      &       0.11$\pm$0.015  &       1.766$\pm$0.176 &       0.76$\pm$0.059  &       143.4$\pm$5.4   \\
2016 Mar 18     &       57465   &       C0      &       1.06$\pm$0.053  &       0.0$\pm$0.0     &       0.003$\pm$0.002 &       0.0$\pm$0.0     \\
        &               &       C1      &       0.869$\pm$0.044 &       0.081$\pm$0.008 &       0.072$\pm$0.011 &       144.6$\pm$2.0   \\
        &               &       J1      &       0.128$\pm$0.012 &       0.278$\pm$0.024 &       0.133$\pm$0.024 &       116.2$\pm$3.8   \\
        &               &       J2      &       0.043$\pm$0.012 &       0.62$\pm$0.085  &       0.255$\pm$0.044 &       133.6$\pm$7.3   \\
        &               &       J3      &       0.047$\pm$0.013 &       1.201$\pm$0.128 &       0.379$\pm$0.052 &       126.5$\pm$5.7   \\
        &               &       J4      &       0.106$\pm$0.015 &       1.817$\pm$0.207 &       0.856$\pm$0.064 &       143.9$\pm$6.2   \\
2016 Apr 22     &       57500   &       C0      &       1.303$\pm$0.065 &       0.0$\pm$0.0     &       0.0$\pm$0.0     &       0.0$\pm$0.0     \\
        &               &       C1      &       0.682$\pm$0.035 &       0.087$\pm$0.008 &       0.074$\pm$0.012 &       146.5$\pm$2.2   \\
        &               &       J1      &       0.112$\pm$0.011 &       0.301$\pm$0.025 &       0.132$\pm$0.025 &       119.7$\pm$3.8   \\
        &               &       J2      &       0.037$\pm$0.012 &       0.663$\pm$0.09  &       0.249$\pm$0.045 &       132.3$\pm$7.2   \\
        &               &       J3      &       0.024$\pm$0.013 &       1.264$\pm$0.105 &       0.226$\pm$0.048 &       127.0$\pm$4.4   \\
        &               &       J4      &       0.123$\pm$0.016 &       1.742$\pm$0.223 &       0.98$\pm$0.066  &       142.1$\pm$6.9   \\
2016 Jun 10     &       57549   &       C0      &       1.746$\pm$0.087 &       0.0$\pm$0.0     &       0.006$\pm$0.003 &       0.0$\pm$0.0     \\
        &               &       C1      &       0.842$\pm$0.043 &       0.088$\pm$0.007 &       0.061$\pm$0.01  &       150.2$\pm$1.5   \\
        &               &       J1      &       0.14$\pm$0.012  &       0.286$\pm$0.033 &       0.193$\pm$0.028 &       118.6$\pm$5.6   \\
        &               &       J2      &       0.034$\pm$0.012 &       0.691$\pm$0.089 &       0.238$\pm$0.044 &       136.4$\pm$6.9   \\
        &               &       J3      &       0.071$\pm$0.014 &       1.36$\pm$0.159  &       0.559$\pm$0.057 &       130.0$\pm$6.3   \\
        &               &       J4      &       0.103$\pm$0.016 &       1.979$\pm$0.291 &       1.119$\pm$0.073 &       150.8$\pm$8.0   \\
2016 Jul 04      &       57573   &       C0      &       1.952$\pm$0.098 &       0.0$\pm$0.0     &       0.003$\pm$0.002 &       0.0$\pm$0.0     \\
        &               &       C1      &       0.647$\pm$0.033 &       0.085$\pm$0.008 &       0.069$\pm$0.012 &       144.3$\pm$2.2   \\
        &               &       J1      &       0.102$\pm$0.011 &       0.336$\pm$0.029 &       0.146$\pm$0.027 &       121.0$\pm$4.1   \\
        &               &       J2      &       0.032$\pm$0.013 &       0.774$\pm$0.122 &       0.298$\pm$0.051 &       129.9$\pm$8.5   \\
        &               &       J3      &       0.032$\pm$0.013 &       1.386$\pm$0.111 &       0.279$\pm$0.049 &       130.5$\pm$4.3   \\
        &               &       J4      &       0.096$\pm$0.015 &       1.821$\pm$0.259 &       0.982$\pm$0.07  &       146.5$\pm$7.7   \\
2016 Jul 31     &       57600   &       C0      &       2.937$\pm$0.147 &       0.0$\pm$0.0     &       0.013$\pm$0.003 &       0.0$\pm$0.0     \\
        &               &       C1      &       0.3$\pm$0.017   &       0.099$\pm$0.009 &       0.052$\pm$0.012 &       132.0$\pm$2.1   \\
        &               &       J1      &       0.104$\pm$0.011 &       0.367$\pm$0.026 &       0.131$\pm$0.025 &       124.2$\pm$3.3   \\
        &               &       J2      &       0.02$\pm$0.013  &       0.791$\pm$0.104 &       0.206$\pm$0.048 &       128.7$\pm$7.1   \\
        &               &       J3      &       0.035$\pm$0.013 &       1.442$\pm$0.098 &       0.258$\pm$0.046 &       132.8$\pm$3.6   \\
        &               &       J4      &       0.071$\pm$0.015 &       1.675$\pm$0.22  &       0.734$\pm$0.065 &       152.6$\pm$7.1   \\
2016 Sep 5	    &	     57636	 &	     C0	     &	     3.576$\pm$0.179 &	     0.0$\pm$0.0	 &	     0.032$\pm$0.005 &	    0.0$\pm$0.0	      \\
 	&	 	&	C1	&	0.231$\pm$0.013	&	0.116$\pm$0.006	&	0.018$\pm$0.008	&	129.8$\pm$0.6	\\
 	&	 	&	J1	&	0.108$\pm$0.011	&	0.345$\pm$0.03	&	0.154$\pm$0.027	&	120.4$\pm$4.1	\\
 	&	 	&	J2	&	0.038$\pm$0.013	&	0.812$\pm$0.118	&	0.316$\pm$0.05	&	130.8$\pm$7.8	\\
 	&	 	&	J3	&	0.037$\pm$0.012	&	1.409$\pm$0.077	&	0.217$\pm$0.042	&	131.0$\pm$2.9	\\
 	&	 	&	J4	&	0.064$\pm$0.014	&	1.617$\pm$0.18	&	0.594$\pm$0.06	&	149.5$\pm$6.0	\\     
2016 Oct 23	&	57684	&	C0	&	2.975$\pm$0.149	&	0.0$\pm$0.0	&	0.066$\pm$0.008	&	0.0$\pm$0.0	\\
 	&	 	&	J1	&	0.084$\pm$0.011	&	0.359$\pm$0.026	&	0.118$\pm$0.025	&	127.4$\pm$3.4	\\
 	&	 	&	J2	&	0.027$\pm$0.013	&	0.94$\pm$0.14	&	0.311$\pm$0.054	&	129.4$\pm$8.1	\\
 	&	 	&	J3	&	0.019$\pm$0.01	&	1.486$\pm$0.024	&	0.051$\pm$0.024	&	131.9$\pm$0.7	\\
 	&	 	&	J4	&	0.067$\pm$0.015	&	1.692$\pm$0.303	&	0.936$\pm$0.075	&	145.0$\pm$9.8	\\     
2016 Nov 28     &       57720   &       C0      &       1.423$\pm$0.071 &       0.0$\pm$0.0     &       0.021$\pm$0.005 &       0.0$\pm$0.0     \\
        &               &       C2      &       0.941$\pm$0.047 &       0.077$\pm$0.007 &       0.057$\pm$0.009 &       133.6$\pm$1.5   \\
        &               &       J1      &       0.066$\pm$0.011 &       0.377$\pm$0.041 &       0.163$\pm$0.031 &       117.9$\pm$5.4   \\
        &               &       J2      &       0.019$\pm$0.014 &       0.764$\pm$0.167 &       0.304$\pm$0.058 &       132.7$\pm$11.8  \\
        &               &       J3      &       0.032$\pm$0.013 &       1.425$\pm$0.128 &       0.315$\pm$0.052 &       127.5$\pm$4.8   \\
        &               &       J4      &       0.066$\pm$0.015 &       1.726$\pm$0.225 &       0.724$\pm$0.066 &       144.6$\pm$7.1   \\
2016 Dec 23	    &	    57745	&	    C0	    &	    1.925$\pm$0.096	&	   0.0$\pm$0.0	    &	    0.019$\pm$0.005	&	    0.0$\pm$0.0	\\
 	&	 	        &	C2          &       1.021$\pm$0.052	&       0.098$\pm$0.007	&       0.065$\pm$0.01	&       140.7$\pm$1.3	\\
 	&	 	        &	J1          &       0.088$\pm$0.011	&       0.415$\pm$0.04	&       0.186$\pm$0.031	&       124.6$\pm$4.8	\\
 	&	 	        &	J3          &       0.011$\pm$0.016	&       1.411$\pm$0.372	&       0.441$\pm$0.081	&       120.3$\pm$14.4	\\
 	&	            &	J4          &       0.135$\pm$0.015	&       1.608$\pm$0.182	&       0.864$\pm$0.06	&       142.7$\pm$6.1	\\   
2017 Jan 14     &       57767   &       C0      &       2.275$\pm$0.114 &       0.0$\pm$0.0     &       0.014$\pm$0.004 &       0.0$\pm$0.0     \\
        &               &       C2      &       0.772$\pm$0.039 &       0.109$\pm$0.008 &       0.076$\pm$0.012 &       141.0$\pm$1.7   \\
        &               &       J1      &       0.079$\pm$0.011 &       0.396$\pm$0.032 &       0.139$\pm$0.028 &       120.6$\pm$3.8   \\
        &               &       J2      &       0.022$\pm$0.014 &       0.815$\pm$0.17  &       0.333$\pm$0.059 &       134.1$\pm$11.3  \\
        &               &       J3      &       0.044$\pm$0.013 &       1.468$\pm$0.094 &       0.282$\pm$0.045 &       129.0$\pm$3.4   \\
        &               &       J4      &       0.1$\pm$0.015   &       1.806$\pm$0.209 &       0.835$\pm$0.064 &       147.3$\pm$6.3   \\
2017 Feb 04      &       57788   &       C0      &       2.251$\pm$0.113 &       0.0$\pm$0.0     &       0.013$\pm$0.004 &       0.0$\pm$0.0     \\
        &               &       C2      &       0.517$\pm$0.027 &       0.13$\pm$0.008  &       0.056$\pm$0.011 &       142.2$\pm$1.3   \\
        &               &       J1      &       0.094$\pm$0.012 &       0.423$\pm$0.041 &       0.197$\pm$0.031 &       124.1$\pm$4.9   \\
        &               &       J3      &       0.06$\pm$0.013  &       1.408$\pm$0.121 &       0.408$\pm$0.051 &       129.2$\pm$4.6   \\
        &               &       J4      &       0.076$\pm$0.015 &       1.881$\pm$0.269 &       0.906$\pm$0.071 &       146.9$\pm$7.8   \\
2017 Mar 19	    &	    57831	&	    C0	    &	    1.702$\pm$0.085	&	    0.0$\pm$0.0	    &	    0.017$\pm$0.004	&	    0.0$\pm$0.0     \\
 	    &	 	        &	    C2	    &	    0.442$\pm$0.023	&	    0.139$\pm$0.009	&	    0.074$\pm$0.013	&	    145.9$\pm$1.8	        \\
 	    &	 	        &	    J1	    &	    0.056$\pm$0.011	&	    0.429$\pm$0.048	&	    0.174$\pm$0.034	&	    124.0$\pm$5.7	        \\
 	    &	 	        &	    J4	    &	    0.1$\pm$0.015	&	    1.572$\pm$0.202	&	    0.812$\pm$0.063	&	     138.2$\pm$7.0	        \\
2017 Apr 16     &       57859   &       C0      &       1.827$\pm$0.092	&	    0.0$\pm$0.0	    &	    0.025$\pm$0.005	&	    0.0$\pm$0.0     \\
        &               &       C2      &       0.314$\pm$0.018	&	    0.158$\pm$0.01	&	    0.074$\pm$0.014	&	    146.0$\pm$2.0   \\
        &               &       J1      &       0.059$\pm$0.011	&	    0.427$\pm$0.05	&	    0.185$\pm$0.034	&	    124.8$\pm$6.0   \\
        &               &       J4      &       0.1$\pm$0.015	&	    1.607$\pm$0.214	&	    0.853$\pm$0.065	&	    139.2$\pm$7.2  \\
2017 May 13     &       57886   &       C0      &       2.423$\pm$0.121 &       0.0$\pm$0.0     &       0.024$\pm$0.005 &       0.0$\pm$0.0     \\
        &               &       C2      &       0.46$\pm$0.024  &       0.167$\pm$0.009	&	    0.073$\pm$0.013	&	    148.4$\pm$1.5   \\
        &               &       J1      &       0.089$\pm$0.012	&	    0.411$\pm$0.061	&	    0.273$\pm$0.038	&	    124.4$\pm$7.7   \\
        &               &       J4      &       0.109$\pm$0.014	&	    1.595$\pm$0.144	&	    0.638$\pm$0.055	&	    133.6$\pm$4.9   \\
2017 Jun 08      &       57912   &       C0      &       2.333$\pm$0.117 &       0.0$\pm$0.0     &       0.041$\pm$0.006 &       0.0$\pm$0.0     \\
        &               &       C2      &       0.345$\pm$0.019 &       0.186$\pm$0.01  &       0.067$\pm$0.013 &       147.1$\pm$1.4   \\
        &               &       J1      &       0.091$\pm$0.012 &       0.385$\pm$0.052 &       0.239$\pm$0.035 &       127.3$\pm$6.9   \\
        &               &       J4      &       0.116$\pm$0.015 &       1.626$\pm$0.168 &       0.751$\pm$0.058 &       137.4$\pm$5.6   \\
2017 Jul 03      &       57937   &       C0      &       2.431$\pm$0.122 &       0.0$\pm$0.0     &       0.03$\pm$0.005  &       0.0$\pm$0.0     \\
        &               &       C2      &       0.372$\pm$0.02  &       0.191$\pm$0.011 &       0.085$\pm$0.015 &       148.5$\pm$1.7   \\
        &               &       J1      &       0.078$\pm$0.012 &       0.407$\pm$0.059 &       0.248$\pm$0.037 &       125.2$\pm$7.5   \\
        &               &       J4      &       0.127$\pm$0.015 &       1.656$\pm$0.176 &       0.815$\pm$0.059 &       139.0$\pm$5.7   \\
2017 Aug 06      &       57971   &       C0      &       2.988$\pm$0.149 &       0.0$\pm$0.0     &       0.024$\pm$0.005 &       0.0$\pm$0.0     \\
        &               &       C2      &       0.301$\pm$0.017 &       0.184$\pm$0.01  &       0.073$\pm$0.014 &       148.7$\pm$1.7   \\
        &               &       J1      &       0.123$\pm$0.013 &       0.413$\pm$0.062 &       0.324$\pm$0.038 &       136.4$\pm$7.7   \\
        &               &       J4      &       0.091$\pm$0.015 &       1.57$\pm$0.185  &       0.721$\pm$0.061 &       133.5$\pm$6.4   \\
2017 Sep 04      &       58000   &       C0      &       3.672$\pm$0.184 &       0.0$\pm$0.0     &       0.037$\pm$0.005 &       0.0$\pm$0.0     \\
        &               &       C2      &       0.223$\pm$0.014 &       0.219$\pm$0.011 &       0.067$\pm$0.015 &       146.2$\pm$1.6   \\
        &               &       J1      &       0.124$\pm$0.013 &       0.347$\pm$0.067 &       0.351$\pm$0.039 &       134.8$\pm$10.1  \\
        &               &       J4      &       0.101$\pm$0.015 &       1.632$\pm$0.174 &       0.719$\pm$0.059 &       138.2$\pm$5.8   \\
2017 Nov 06      &       58063   &       C0      &       3.464$\pm$0.173 &       0.0$\pm$0.0     &       0.048$\pm$0.006 &       0.0$\pm$0.0     \\
        &               &       C2      &       0.113$\pm$0.011 &       0.273$\pm$0.018 &       0.09$\pm$0.02   &       145.0$\pm$2.7   \\
        &               &       J1      &       0.091$\pm$0.013 &       0.422$\pm$0.068 &       0.305$\pm$0.04  &       129.8$\pm$8.5   \\
        &               &       J4      &       0.077$\pm$0.015 &       1.62$\pm$0.203  &       0.719$\pm$0.063 &       136.0$\pm$6.8   \\
2018 Mar 10	&	58187	&	C0	&	1.832$\pm$0.092	&	0.0$\pm$0.0	&	0.011$\pm$0.004	&	0.0$\pm$0.0	\\
 	&	 	&	C3	&	1.256$\pm$0.063	&	0.042$\pm$0.009	&	0.105$\pm$0.012	&	89.1$\pm$5.0	\\
 	&	 	&	J1	&	0.117$\pm$0.013	&	0.415$\pm$0.055	&	0.287$\pm$0.036	&	136.3$\pm$6.9	\\
 	&	 	&	J4	&	0.049$\pm$0.014	&	1.506$\pm$0.187	&	0.534$\pm$0.061	&	135.1$\pm$6.7	\\      
2018 Apr 19	&	58227	&	C0	&	1.695$\pm$0.085	&	0.0$\pm$0.0	&	0.011$\pm$0.004	&	0.0$\pm$0.0	\\
 	&	 	&	C3	&	0.844$\pm$0.043	&	0.062$\pm$0.009	&	0.102$\pm$0.013	&	111.5$\pm$4.0	\\
 	&	 	&	J1	&	0.108$\pm$0.013	&	0.438$\pm$0.066	&	0.322$\pm$0.039	&	131.1$\pm$7.9	\\
 	&	 	&	J4	&	0.063$\pm$0.014	&	1.6$\pm$0.187	&	0.606$\pm$0.061	&	135.8$\pm$6.3	\\        
2018 May 11     &       58249   &       C0      &       1.344$\pm$0.067 &       0.0$\pm$0.0     &       0.005$\pm$0.003 &       0.0$\pm$0.0     \\
        &               &       C3      &       0.704$\pm$0.036 &       0.077$\pm$0.01  &       0.101$\pm$0.014 &       126.0$\pm$3.6   \\
        &               &       J1      &       0.1$\pm$0.013   &       0.44$\pm$0.067  &       0.313$\pm$0.039 &       134.8$\pm$7.9   \\
        &               &       J4      &       0.072$\pm$0.014 &       1.617$\pm$0.167 &       0.587$\pm$0.058 &       137.1$\pm$5.6   \\
2018 Jun 16     &       58285   &       C0      &       1.421$\pm$0.071 &       0.0$\pm$0.0     &       0.011$\pm$0.004 &       0.0$\pm$0.0     \\
        &               &       C3      &       0.418$\pm$0.022 &       0.107$\pm$0.013 &       0.113$\pm$0.016 &       129.2$\pm$4.1   \\
        &               &       J1      &       0.08$\pm$0.012  &       0.482$\pm$0.07  &       0.293$\pm$0.04  &       131.2$\pm$7.6   \\
        &               &       J4      &       0.058$\pm$0.014 &       1.585$\pm$0.168 &       0.531$\pm$0.058 &       134.4$\pm$5.7   \\
2018 Jul 16     &       58315   &       C0      &       2.396$\pm$0.12  &       0.0$\pm$0.0     &       0.033$\pm$0.006 &       0.0$\pm$0.0     \\
        &               &       C3      &       0.337$\pm$0.019 &       0.126$\pm$0.014 &       0.119$\pm$0.018 &       127.9$\pm$4.2   \\
        &               &       J1      &       0.11$\pm$0.013  &       0.441$\pm$0.07  &       0.345$\pm$0.04  &       137.5$\pm$8.4   \\
        &               &       J4      &       0.102$\pm$0.015 &       1.704$\pm$0.194 &       0.795$\pm$0.062 &       133.9$\pm$6.2   \\
2018 Aug 26     &       58356   &       C0      &       1.696$\pm$0.085 &       0.0$\pm$0.0     &       0.053$\pm$0.008 &       0.0$\pm$0.0     \\
        &               &       C3      &       0.077$\pm$0.009 &       0.208$\pm$0.01  &       0.033$\pm$0.014 &       118.0$\pm$1.3   \\
        &               &       J1      &       0.049$\pm$0.013 &       0.534$\pm$0.09  &       0.286$\pm$0.045 &       124.9$\pm$9.0   \\
        &               &       J4      &       0.029$\pm$0.014 &       1.559$\pm$0.196 &       0.427$\pm$0.062 &       136.6$\pm$6.8   \\
\noalign{\smallskip}
\end{longtable}
\tablefoot{The columns are (1) observing date in {year month day format}, (2) observing date in modified Julian date, (3) {identification} of the jet component, (4) flux density in Jy, (5) relative distance from the core in mas, (6) size of the jet component in mas, and (7) relative position angle with respect to the core in {degrees}.}
\end{appendix}